\documentclass[aps, a4paper,superscriptaddress,12pt,preprintnumbers,floatfix,nofootinbib,amsmath,amssymb]{revtex4}
\usepackage{amstext,amssymb}
\usepackage{tikz}
\usepackage{tikz-feynman}
\usepackage{tikz-feynman,contour}
\usepackage{amsmath}
\usepackage{graphicx}
\usepackage[hyperfootnotes=false]{hyperref}
\usepackage{xspace}
\usepackage{color}
\usepackage{units}
\usepackage{subfigure}
\usepackage{slashed} 

\newcommand{\bea}{\begin{eqnarray}}
\newcommand{\eea}{\end{eqnarray}}
\newcommand{\nn}{\nonumber}

\global\long\def\d{\partial}

\def\s1{\hat s}

\def\U1mt{U(1)_{L_\mu-L_\tau}}

\def\SO10{\text{SO}(10)}

\def \epsilon {\varepsilon} 

\usepackage{longtable}
\usepackage{needspace}
\def\SO10{\text{SO}(10)}

\newcommand{\red}[1]{\textcolor{red}{#1}}

\begin{document}
\title{\hspace*{-1.0cm} Neutrino magnetic moment and inert doublet dark matter in a Type-III radiative scenario} 
\author{Shivaramakrishna \surname{Singirala}}
\email{krishnas542@gmail.com}
\affiliation{School of Physics, University of Hyderabad, Hyderabad 500046, India}
\author{Dinesh Kumar \surname{Singha}}
\email{dinesh.sin.187@gmail.com}
\affiliation{School of Physics, University of Hyderabad, Hyderabad 500046, India}
\author{Rukmani \surname{Mohanta}}
\email{rmsp@uohyd.ac.in}
\affiliation{School of Physics, University of Hyderabad, Hyderabad 500046, India}

\begin{abstract}
\vspace*{1cm}

\end{abstract}
\maketitle
We narrate dark matter, neutrino magnetic moment and mass in a Type-III radiative scenario. The Standard Model is enriched with three vector-like fermion triplets and two inert doublets to provide a suitable platform for the above phenomenological aspects.  The inert scalars contribute to total relic density of dark matter in the Universe. Neutrino aspects are realized at one-loop with magnetic moment obtained through charged scalars, while neutrino mass gets contribution from charged and neutral scalars. Taking inert scalars up to $2$ TeV and triplet fermion in few hundred TeV range, we obtain a common parameter space, compatible with experimental limits associated with both neutrino and dark matter sectors. Using a specific region for transition magnetic moment (${\cal O} (10^{-11}\mu_B$)), we explain the excess recoil events, reported by the XENON1T collaboration. Finally, we demonstrate that the model is able to provide neutrino magnetic moments in a wide range from $10^{-12}\mu_B$ to $10^{-10}\mu_B$, meeting the bounds of various experiments such as Super-K, TEXONO, Borexino and XENONnT.
\newpage
\section{Introduction}
The fascinating description of elementary particle physics is elegantly portrayed by the Standard Model (SM) in the low energy regime. The locally gauge invariant Lagrangian is able to describe how the interactions proceed at the most fundamental level.  This gauge theory has provided a pathway to understand the behavior of nature at very tiny length scale and serves as a theoretical torch for exploring  several unknown things beyond, a never ending tale of theorists and experimentalists to unravel the mysteries of the Universe. A small chunk of puzzles include neutrino masses and mixing \cite{Bilenky:1999ge,Mohapatra:1979ia,Schechter:1980gr,Babu:1992ia,Super-Kamiokande:2005wtt,SNO:2002tuh,Super-Kamiokande:2016yck,T2K:2019efw,DayaBay:2012fng,DoubleChooz:2011ymz}, nature and identity of dark matter (DM) \cite{Zwicky:1937zza,PhysRev.43.147,Bertone:2004pz,Mambrini:2015vna}, matter anti-matter asymmetry \cite{Sakharov:1967dj,Kolb:1979qa,Fukugita:1986hr,Fritzsch:1974nn} and the recently observed  anomalies in the flavor sector \cite{Bifani:2018zmi}.

Several neutrino experiments have unambiguously proved that oscillation of neutrino flavor occurs during propagation and neutrinos possess small but non-zero masses.  With this extra degree of freedom, many new possibilities are opened up, and  one such possibility amongst them is that neutrinos can possess electromagnetic properties like electric and magnetic dipole moments. Solar, accelerator and reactor experiments help us in the direct measurement of magnetic moments and eventually put the  limits on them. One possible mode of measurement  involves the study of neutrino/anti-neutrino electron scattering at low energy limits. In minimally extended SM, one can have neutrino magnetic moment ($\nu$MM) of the order $10^{-19}$ $\mu_{B}$ for Dirac type of neutrinos. However, these values are beyond the  sensitivity reach of any  experimental measurement. On the other hand, for Majorana type neutrinos,  one can have a very high transition magnetic moment, fitting the experimental observations. That's why the study of $\nu$MM becomes important for the distinction of Dirac and Majorana type of neutrinos.

In recent past, XENON collaboration performed a search for new physics with its $1$ ton detector and reported an excess of events over the known backgrounds in the recoil energy range $1-7$ keV, peaked around $2.5$ keV \cite{XENON:2020rca}. It turns out that such excess can be explained with large transition magnetic moment of neutrinos. With new data from its successor XENONnT \cite{XENON:2022ltv}, no visibly bold excess events were seen in the low energy region creating an anomalous situation between these two experiments. The collaboration is suspecting this excess in XENON1T was due to uncounted tritium whose presence or absence they can't corroborate. In this scenario, we cannot completely ignore the  possible implication of new physics effects at XENON1T and that is why it is very interesting to explore such possibilities. Several works explaining this excess and neutrino electromagnetic properties can be found in the literature \cite{Miranda:2020kwy, Li:2022bqr, Miranda:2021kre, Babu:2021jnu, Brdar:2020quo, Khan:2022bel, Khan:2022akj, Jeong:2021ivd,Alok:2022pdn,Dror:2020czw} .

Zwicky made a proposal in 1933 for the existence of dark matter through observations of spiral galaxy rotation curves, however the physics of this mysterious particle is still unsettled. Freeze-out scenario has been the one that has fascinated theoretical physicists, a paradigm that is able to provide proper relic density as per Planck satellite by a weakly interacting massive particle (WIMP). Now, we raise a question, whether a dark matter particle running in the loop, forming an electromagnetic vertex can provide neutrino magnetic moment. With this view point, we provide a simple model that can accommodate non-zero magnetic moment for neutrino and also discuss dark matter phenomenology in a correlative manner. We check the sensitivity of $\nu$MM with XENONnT \cite{XENON:2022ltv}, Borexino \cite{Borexino:2017fbd}, TEXONO \cite{TEXONO:2006xds}, Super-K \cite{Super-Kamiokande:2004wqk} and white dwarfs \cite{MillerBertolami:2014oki} and also explain the excess in electron recoil events reported at XENON1T \cite{XENON:2020rca}. 

The paper is organized as follows. In section-II, we describe the model along with the particle content and interaction terms to address neutrino magnetic moment, neutrino mass and dark matter. The mass spectrum of scalar sector due to mixing is also discussed in this section. Section-III  narrates neutrino magnetic moment and neutrino masses at one-loop. Section-IV describes the   dark matter relic density and its detection prospects. Section-V provides the detailed analysis, showing common parameter space to obtain observables related to the aspects of neutrino and dark matter sectors. We also emphasize more specific constraints on Yukawa couplings from current neutrino oscillation data. Section-VI gives the signature of magnetic moment in the light of electron recoil event excess at XENON1T and also the overall obtained range of magnetic moment in the concerned model. Finally, concluding remarks are provided in section-VII.

\section{Model description}
To address the neutrino mass,  magnetic moment and dark matter in a common platform, we extend the SM framework with three vector-like fermion triplets $\Sigma_k$, with $k=1,2,3$ and two inert scalar doublets $\eta_j$, with $j=1,2$. We impose an additional $Z_2$ symmetry to realize neutrino phenomenology at one-loop and also for the stability of the dark matter candidate. The particle content along with their charges are displayed in Table. \ref{typeiii_model}.

\begin{table}[htb]
\caption{Fields and their charges in the present model.}
\begin{center}
\begin{tabular}{|c|c|c|c|c|}
	\hline
			& Field	& $ SU(3)_C \times SU(2)_L\times U(1)_Y$ & $Z_2$\\
	\hline
	\hline
	Leptons	& $\ell_{L} = (\nu, e)^T_L$	& $(\textbf{1},\textbf{2},~  -1/2)$ & $+$\\
			& $e_R$							& $(\textbf{1},\textbf{1},~  -1)$	& $+$\\
	&  $\Sigma_{k(L,R)}$ & $(\textbf{1},\textbf{3},0)$ & $-$ \\
\hline	
	Scalars	& $H$							& $(\textbf{1},\textbf{2},~ 1/2)$	&  $+$\\ 
					& $\eta_j$							& $(\textbf{1},\textbf{2},~ 1/2)$	& $-$\\
			\hline
	\hline
\end{tabular}
\label{typeiii_model}
\end{center}
\end{table}
The $SU(2)_L$ triplet $\Sigma_{L,R} = \left(\Sigma^1,\Sigma^2,\Sigma^3 \right)_{L,R}^T$ can be expressed in the fundamental representation as
\begin{equation}
\Sigma_{L,R} = \frac{\sigma^a \Sigma^a_{L,R}}{\sqrt{2}} = \begin{pmatrix}
		 \Sigma^0_{L,R}/\sqrt{2}	& \Sigma^+_{L,R}	\\
		 \Sigma^-_{L,R}	& -\Sigma^0_{L,R}/\sqrt{2}	\\
	\end{pmatrix}.
\end{equation}
Here, $\sigma^a$'s represent Pauli matrices and $\Sigma^0_{L,R} = \Sigma^3_{L,R}$, $\Sigma^{\pm}_{L,R} = \left(\Sigma^1_{L,R} \mp \Sigma^2_{L,R} \right)/\sqrt{2}$. The Lagrangian terms of the model is given by
\begin{eqnarray}
\mathcal{L}_{\Sigma} &=& y^\prime_{\alpha k} \overline{\ell_{\alpha L}}\Sigma_{kR} \tilde{\eta}_{j} + y_{\alpha k} \overline{\ell^c_{\alpha L}}i\sigma_2 \Sigma_{kL} \eta_{j}  + \frac{i}{2}{\rm Tr}[\overline{\Sigma}\gamma^\mu D_\mu\Sigma] - \frac{1}{2}{\rm Tr}[\overline{\Sigma}M_\Sigma\Sigma] + {\rm h.c.}
\end{eqnarray}
In the above, $\Sigma^{+,0} = \Sigma_L^{+,0} + \Sigma_R^{+,0}$   and $\Sigma = (\Sigma_1,\Sigma_2,\Sigma_3)^T$. The covariant derivative for $\Sigma$ is given by
\begin{eqnarray}
&& D_\mu \Sigma= \d_\mu \Sigma + ig \left[\sum_{a=1}^3 \frac{\sigma^a}{2}W^a_\mu, \Sigma\right].
\end{eqnarray}
 The  Lagrangian for the scalar sector takes the form 
 \begin{align}
\mathcal{L}_{\rm scalar} &= \left|\left(\d_\mu  + \frac{i}{2} g~  \sigma^a W^a_\mu  + \frac{i}{2}g^\prime B_\mu\right) \eta_1\right|^2 +\left|\left(\d_\mu  + \frac{i}{2} g~  \sigma^a W^a_\mu  + \frac{i}{2}g^\prime B_\mu\right) \eta_2\right|^2 - V(H,\eta_1, \eta_2),
\end{align}
where, the inert doublets are denoted by $\eta_j = \begin{pmatrix}
		 \eta_j^+		\\
		 \eta_j^0	\\
	\end{pmatrix}$, with $\eta_j^0 = \displaystyle{\frac{\eta^R_j+i\eta^I_j}{\sqrt{2}}}$ and the scalar potential is expressed as \cite{Keus:2014jha,Keus:2014isa}
\begin{align}
V(H, \eta_1, \eta_2) &=  \mu^2_H  H^\dagger H +  \mu^2_1 \eta_1^\dagger \eta_1 + \mu^2_2 \eta_2^\dagger \eta_2 +\mu^2_{12} (\eta_1^\dagger \eta_2 + {\rm h.c}) + \lambda_H (H^\dagger H)^2 + \lambda_1 (\eta_1^\dagger \eta_1)^2 \nn\\   
      &+ \lambda_2 (\eta_2^\dagger \eta_2)^2+ \lambda_{12} (\eta_1^\dagger \eta_1) (\eta_2^\dagger \eta_2) + \lambda^\prime_{12} (\eta_1^\dagger \eta_2) (\eta_2^\dagger \eta_1) + \frac{\lambda^{\prime\prime}_{12}}{2} \left[(\eta_1^\dagger \eta_2)^2 + {\rm h.c.}\right] 
       \nonumber\\ &+ \sum_{j=1,2} \left(\lambda_{Hj} (H^\dagger H)(\eta_j^\dagger \eta_j)  
    +  \lambda^\prime_{Hj} (H^\dagger \eta_j)(\eta_j^\dagger H) + \frac{\lambda^{\prime\prime}_{Hj}}{2} \left[(H^\dagger \eta_j)^2 + {\rm h.c}\right]\right).    
\label{scalarpot}
\end{align} 
\newline
\subsection{Copositive criteria}
The above potential has a stable vacuum if \cite{Keus:2014jha}
\begin{eqnarray}
    && \lambda_H,\lambda_1, \lambda_2 > 0,\nn\\
    && \lambda_{H1} + \lambda^\prime_{H1} + 2\sqrt{\lambda_H \lambda_1} > 0,\nn\\
    && \lambda_{H2} + \lambda^\prime_{H2} + 2\sqrt{\lambda_H \lambda_2} > 0,\nn\\
    && \lambda_{12} + \lambda^\prime_{12} + 2\sqrt{\lambda_1 \lambda_2} > 0,\nn\\
    && |\lambda^{\prime\prime}_{H1}|, |\lambda^{\prime\prime}_{H2}|, |\lambda^{\prime\prime}_{12}| < |\lambda_{H}|, |\lambda_{1}| , |\lambda_{2}|, |\lambda_{H1}|,|\lambda^\prime_{H1}|, |\lambda_{H2}|, |\lambda^\prime_{H2}|, |\lambda_{12}|,|\lambda^\prime_{12}|.
\end{eqnarray}

\subsection{Mass spectrum}
The mass matrices of the charged and neural scalar components are given by
\begin{align}
{\mathcal M}^2_C =
	\begin{pmatrix}
		 \Lambda_{C1}	& \mu_{12}	\\
		 \mu_{12}	& \Lambda_{C2}	\\
	\end{pmatrix}, \quad
{\mathcal M}^2_R =
	\begin{pmatrix}
		 \Lambda_{R1}	& \mu_{12}	\\
		 \mu_{12}	& \Lambda_{R2}	\\
	\end{pmatrix}, \quad
{\mathcal M}^2_I =
	\begin{pmatrix}
		 \Lambda_{I1}	& \mu_{12}	\\
		 \mu_{12}	& \Lambda_{I2}	\\
	\end{pmatrix}.
\end{align}
Here, 
\begin{eqnarray}
&&\Lambda_{Cj} = \mu_{j}^2  + \frac{ \lambda_{ Hj}}{2} v^2,\nn\\
&&\Lambda_{Rj} = \mu_{j}^2  +  \left(\lambda_{Hj} + \lambda'_{Hj} + \lambda''_{Hj}\right)\frac{v^2}{2},\nn\\
&&\Lambda_{Ij} = \mu_{j}^2  +  \left(\lambda_{Hj} + \lambda'_{Hj} - \lambda''_{Hj}\right)\frac{v^2}{2}.
\end{eqnarray}
One can diagonalize the above mass matrices using
$
U_\theta
	=
	\begin{pmatrix}
		 \cos{\theta}	& \sin{\theta}	\\
		 -\sin{\theta}	& \cos{\theta}	\\
	\end{pmatrix}
$ as
\begin{eqnarray}
&& U_{\theta_C}^T {\mathcal M}^2_{C} U_{\theta_C} = {\rm{diag}}(M^2_{{C1}},M^2_{{C2}}) \quad {\rm with} \quad
\theta_C = \tan^{-1}\left[\frac{2\mu^2_{12}}{\Lambda_{C2}-\Lambda_{C1}}\right],\nn\\
&& U_{\theta_R}^T {\mathcal M}^2_{R} U_{\theta_R} = {\rm{diag}}(M^2_{{R1}},M^2_{{R2}}) \quad {\rm with} \quad
\theta_R = \tan^{-1}\left[\frac{2\mu^2_{12}}{\Lambda_{R2}-\Lambda_{R1}}\right],\nn\\
&& U_{\theta_I}^T {\mathcal M}^2_{I} U_{\theta_I} = {\rm{diag}}(M^2_{{I1}},M^2_{{I2}}) \quad {\rm with} \quad
\theta_I = \tan^{-1}\left[\frac{2\mu^2_{12}}{\Lambda_{I2}-\Lambda_{I1}}\right].
\end{eqnarray}
The flavor and mass eigen states can be related as
\begin{align}
	\begin{pmatrix}
		 \eta^+_1\\
		 \eta^+_2\\
	\end{pmatrix} =
	U_{\theta_C}
		\begin{pmatrix}
		 \phi^+_1\\
		 \phi^+_2\\
	\end{pmatrix}, \quad
	\begin{pmatrix}
		 \eta^R_1\\
		 \eta^R_2\\
	\end{pmatrix} =
	U_{\theta_R}		\begin{pmatrix}
		 \phi^R_1\\
		 \phi^R_2\\
	\end{pmatrix}, \quad
	\begin{pmatrix}
		 \eta^I_1\\
		 \eta^I_2\\
	\end{pmatrix} =
	U_{\theta_I}		\begin{pmatrix}
		 \phi^I_1\\
		 \phi^I_2\\
	\end{pmatrix}.
\end{align}
The invisible decays of $Z$ and $W^\pm$ at LEP, limit the masses of inert scalars as \cite{Cao:2007rm,Lundstrom:2008ai}
\begin{eqnarray}
&&    M_{Ci} > M_Z/2, \quad M_{Ri} + M_{Ii} > M_Z, \quad M_{Ci} + M_{Ri, Ii} > M_{W}.
\end{eqnarray}
Moving on to fermion sector, electroweak radiative corrections provide a mass splitting of $166$ MeV \cite{Cirelli:2005uq} between the charged and neutral component of triplet. We work in the high scale regime, this small splitting does not effect the phenomenology. 
\section{Neutrino phenomenology}
\subsection{Neutrino Magnetic moment}
\begin{figure}
  \centering
  \begin{tikzpicture}
    \begin{feynman}
      \vertex (a) at (-2,-2);
      \vertex (b) at ( 0,2) ;
      \vertex (c) at (2, -2) ;
      \vertex[blob] (m) at ( 0, 0) {\contour{white}{\red{$\mathbf{\Gamma_{\mu}}$}}};
      \diagram* {
        (a) -- [fermion, very thick, edge label=$\psi (p)$] (m) -- [fermion, very thick, edge label=$\psi (p')$] (c),
        (b) -- [boson, very thick, edge label'=$A_{\mu} (q)$] (m),
      };
    \end{feynman}
  \end{tikzpicture}
  \caption{Effective electromagnetic vertex, where $q = p-p^\prime$.  }
  \label{emvtx_feyn}
\end{figure}
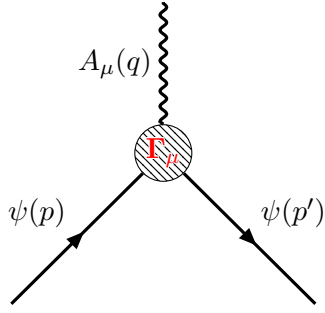

Though neutrino is electrically neutral, it can have electromagnetic interaction at loop level, as shown in Fig. \ref{emvtx_feyn}, where $\psi(p)$ and $\psi(p')$ denote the incoming and outgoing neutrino states. The effective interaction Lagrangian takes the form \cite{Xing:2011zza}
\begin{equation}
\mathcal{L}_{\rm EM} =  \overline{\psi}\Gamma_\mu \psi A^\mu\;.
\end{equation}
In the above, the electromagnetic vertex function varies with the type of neutrinos, i.e., Dirac or Majorana. In case of Dirac neutrino, $\Gamma_\mu$ takes the form
\begin{equation}
\Gamma_\mu (p,p^\prime) =  f_Q(q^2)\gamma_\mu + if_M(q^2)\sigma_{\mu\nu}q^\nu + f_E(q^2) \sigma_{\mu\nu}q^\nu \gamma_5 + f_A(q^2) (q^2\gamma_\mu - q_\mu\slashed{q})\gamma_5\;,
\end{equation}
where $f_Q(q^2),~f_M(q^2),~ f_E(q^2)$ and $f_A(q^2)$ represent the form factors of charge, magnetic dipole, electric dipole and anapole respectively.
%
%

In the non-relativisitic regime, $f_Q(0)=Q$ stands for the charge, $f_M(0)=\mu$ represents magnetic dipole moment, $f_E(0)$ denotes electric dipole moment and $f_A(0)$ stands for the Zeldovich anapole moment of the particle. All the four form factors remain finite in Dirac type neutrino. For Majorana case, using the property of charge conjugation $\psi^c =  C\overline{\psi}^T$, we get
\begin{equation}
\overline{\psi}\Gamma_\mu \psi = \overline{\psi^c}\Gamma_\mu \psi^c = \overline{\psi}C\Gamma_\mu^T C^{-1} \psi.
\end{equation}
Since $C\gamma_\mu^T C^{-1}=-\gamma_\mu$, $C(\gamma_\mu \gamma_5)^T C^{-1} = \gamma_\mu \gamma_5$, $C\sigma_{\mu\nu}^T C^{-1} = -\sigma_{\mu\nu}$ and $C(\sigma_{\mu\nu}\gamma_5)^T C^{-1} = -\sigma_{\mu\nu}\gamma_5$, we obtain
\begin{equation}
\Gamma_\mu (p,p^\prime) =  -f_Q(q^2)\gamma_\mu - if_M(q^2)\sigma_{\mu\nu}q^\nu - f_E(q^2) \sigma_{\mu\nu}q^\nu \gamma_5 + f_A(q^2) (q^2\gamma_\mu - q_\mu\slashed{q})\gamma_5\;,
\end{equation}
which results $f_Q(q^2) = f_M(q^2) = f_E(q^2) = 0$ for a Majorana neutrino. However, if the electromagnetic current is between two different neutrino flavors in the initial and final states i.e., $\overline{\psi_i}\Gamma_\mu \psi_j A^\mu$ with $i\neq j$, Majorana neutrinos can have non-zero transition dipole moments. 

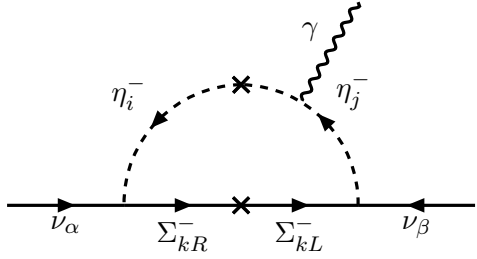
\begin{figure}
    \begin{tikzpicture}
        \begin{feynman}
            \vertex (a);
            \vertex[right=4em of a] (b);
            \vertex[right=4em of b] (c);
            \vertex[right=4em of c] (e);
            \vertex[right=4em of e] (f);
            \vertex at ($(b)!0.5!(e)!1.6cm!+90:(e)$) (d);
            \vertex at ($(b)!0.5!(e)!1.6cm!+60:(e)$) (g);
            \vertex[above=7em of e] (h);

            \diagram* {
      (a) -- [fermion, very thick, edge label'=$\nu_{\alpha}$] (b),
      (e) -- [charged scalar, very thick, quarter right, edge label'=$\eta^-_j$]
      (d) -- [charged scalar, very thick, quarter right, insertion=0, edge label'=$\eta^-_i$] (b),
      (b) -- [fermion, very thick, edge label'=$\Sigma^-_{kR}$] (c),
      (c) -- [fermion, very thick, insertion=0, edge label'=$\Sigma^-_{kL}$] (e),
      (f) -- [fermion, very thick, edge label=$\nu_{\beta}$] (e),
      (g) -- [boson, very thick, edge label=$\gamma$] (h),
     
    };
        \end{feynman}
    \end{tikzpicture}
    \caption{One-loop Feynman diagram for transition magnetic moment.}
    \label{numag_feyn}
\end{figure}

In the present model, the magnetic moment arises from one-loop diagram shown in Fig. \ref{numag_feyn}, and the expression of corresponding contribution takes the form \cite{Babu:2020ivd}
\begin{eqnarray}
&&(\mu_{\nu})_{\alpha \beta} = \sum^3_{k=1}\frac{({Y^2})_{\alpha \beta}} {8\pi^{2}}M_{\Sigma^+_k}   \bigg[(1+\sin 2\theta_C)\frac{1}{M^{2}_{C2}} \left(\ln \left[\frac{M^{2}_{C2}}{M^2_{\Sigma^+_k}}\right]-1\right) \nonumber\\
&&~~~~~~~~~~~~~~+ (1-\sin 2\theta_C)\frac{1}{M^{2}_{C1}} \left(\ln \left[\frac{M^{2}_{C1}}{M^2_{\Sigma^+_k}}\right]-1\right) \bigg],
\label{mag_eqn}
\end{eqnarray}
where $y = y^\prime = Y$ and $({Y^2})_{\alpha \beta}= {Y}_{\alpha k} {Y}_{k \beta}^T$.
\subsection{Neutrino mass}
In the present model, contribution to neutrino mass  can arise at one-loop from two diagrams, one with charged scalars  and fermion triplet in the loop while the other with neutral scalars and fermion triplets. The relevant diagrams are provided in Fig. \ref{numass_feyn} and the corresponding contribution takes the form \cite{Chen:2020ark, Lineros:2020eit, Avila:2019hhv}
\begin{eqnarray}
&&({\cal M}_{\nu})_{\alpha\beta} = \sum_{k=1}^3 \frac{({Y^2})_{\alpha \beta}}{32 \pi^{2}}M_{\Sigma^+_k}
 \Bigg[(1+\sin2\theta_C)\frac{M_{C2}^2}{M_{\Sigma^+_k}^{2}-M_{C2}^2}\ln\left(\frac{M^2_{\Sigma^+_k}}{M^2_{C2}}\right)\nonumber\\
 &&~~~~~~~~~~~~~~~~~~~~~~~~~~~~~~~~+ (1-\sin2\theta_C)\frac{M_{C1}^2}{M_{\Sigma^+_k}^{2} -  M_{C1}^2}\ln\left(\frac{M^2_{\Sigma^+_k}}{M^2_{C1}}\right)  \Bigg]  \nonumber\\
&&~~~~~~~~+\sum_{k=1}^3 \frac{{(Y^2)}_{\alpha \beta}}{32 \pi^{2}} M_{\Sigma^0_k}
 \Bigg[(1+\sin2\theta_R)\frac{M_{R2}^2}{M_{\Sigma^0_k}^{2}-M_{R2}^2}\ln\left(\frac{M^2_{\Sigma^0_k}}{M^2_{R2}}\right) \nonumber\\
 &&~~~~~~~~~~~~~~~~~~~~~~~~~~~~~~~+ (1-\sin2\theta_R)\frac{M_{R1}^2}{M_{\Sigma^0_k}^{2} -  M_{R1}^2}\ln\left(\frac{M^2_{\Sigma^0_k}}{M^2_{R1}}\right)  \Bigg]  \nonumber\\
 &&~~~~~~~~-\sum_{k=1}^3 \frac{{(Y^2})_{\alpha \beta}}{32 \pi^{2}} M_{\Sigma^0_k}
 \Bigg[(1+\sin2\theta_I)\frac{M_{I2}^2}{M_{\Sigma^0_k}^{2}-M_{I2}^2}\ln\left(\frac{M^2_{\Sigma^0_k}}{M^2_{I2}}\right)\nonumber\\
 &&~~~~~~~~~~~~~~~~~~~~~~~~~~~~~~~~+ (1-\sin2\theta_I)\frac{M_{I1}^2}{M_{\Sigma^0_k}^{2} -  M_{I1}^2}\ln\left(\frac{M^2_{\Sigma^0_k}}{M^2_{I1}}\right)  \Bigg].
\label{mass_eqn}
\end{eqnarray}
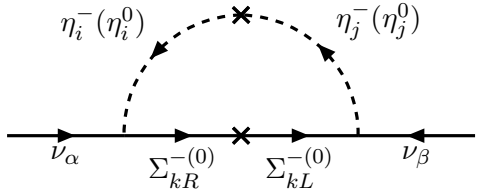
\begin{figure}
    \begin{tikzpicture}
        \begin{feynman}
            \vertex (a);
            \vertex[right=4em of a] (b);
            \vertex[right=4em of b] (c);
            \vertex[right=4em of c] (e);
            \vertex[right=4em of e] (f);
            \vertex at ($(b)!0.5!(e)!1.6cm!+90:(e)$) (d);
            \vertex at ($(b)!0.5!(e)!1.6cm!+60:(e)$) (g);
            \vertex[above=7em of e] (h);

            \diagram* {
      (a) -- [fermion, very thick, edge label'=$\nu_{\alpha}$] (b),
      (e) -- [charged scalar, very thick, quarter right, edge label'=$\eta^-_j (\eta^0_j)$]
      (d) -- [charged scalar, very thick, quarter right, insertion=0, edge label'=$\eta^-_i (\eta^0_i)$] (b),
      (b) -- [fermion, very thick, edge label'=$\Sigma^{-(0)}_{kR}$] (c),
      (c) -- [fermion, very thick, insertion=0, edge label'=$\Sigma^{-(0)}_{kL}$] (e),
      (f) -- [fermion, very thick, edge label=$\nu_{\beta}$] (e),
     
    };
        \end{feynman}
    \end{tikzpicture}
    \caption{One-loop diagram that generates light neutrino mass.}
    \label{numass_feyn}
\end{figure}
\section{Dark Matter phenomenology}
\subsection{Relic density}
The model provides scalar dark matter candidates and we study their phenomenology for dark matter mass up to $2$ TeV range. All the inert scalar components contribute to the dark matter density of the Universe through annihilations and co-annihilations. With the mediation of scalar Higgs, $\phi^R_i \phi^R_j$ can annihilate to $f\bar{f}, ~W^+W^-,ZZ, ~hh$ and via $Z$ boson, $\phi^R_i \phi^I_j$ can co-annihilate to $f\bar{f}, ~W^+W^-,~ Zh$. The charged and neutral components can co-annihilate to $f^\prime \overline{f^{\prime\prime}}, A W^\pm, Z W^\pm, h W^\pm$ through $W^\pm$. Here, $f^\prime = u,c,t,\nu_e,\nu_\mu, \nu_\tau$ and $f^{\prime\prime} = d,s,b,e,\mu,\tau$ \cite{LopezHonorez:2006gr, Barbieri:2006dq, Dolle:2009fn}. The abundance of dark matter can be computed by
\begin{equation}
\label{eq:relicdensity}
\Omega h^2 = \frac{1.07 \times 10^{9} ~{\rm{GeV}}^{-1}}{  M_{\rm{Pl}}\; {g_\ast}^{1/2}}\frac{1}{J(x_f)}\;,
\end{equation}
where, $M_{\rm{Pl}}=1.22 \times 10^{19} ~\rm{GeV}$ and $g_\ast = 106.75$ denote the Planck mass and total number of effective relativistic degrees of freedom respectively. The function $J$ is
\begin{equation}
J(x_f)=\int_{x_f}^{\infty} \frac{ \langle \sigma v \rangle (x)}{x^2} dx.
\end{equation}
In the above, the thermally averaged cross section $\langle \sigma v \rangle$ reads  as
\begin{equation}
 \langle\sigma v\rangle (x) = \frac{x}{8 M_{\rm DM}^5 K_2^2(x)} \int_{4 M_{\rm DM}^2}^\infty \hat{\sigma} \times ( s - 4 M_{\rm DM}^2) \ \sqrt{s} \ K_1 \left(\frac{x \sqrt{s}}{M_{\rm DM}}\right) ds.
\end{equation}
Here $K_1$, $K_2$ are the modified Bessel functions, $x = M_{\rm DM}/T$, with $T$ being the temperature, $M_{\rm  DM}$ is dark matter mass, $\hat  \sigma$ is the dark matter cross section and $x_f$ stands for the freeze-out parameter. 
\subsection{Direct searches}
Moving to direct searches, the scalar dark matter can scatter off the nucleus via the Higgs and the $Z$ boson. Mass splitting between real and imaginary components above $100$ KeV can forbid gauge kinematics \cite{Dolle:2009fn}. Thus the DM-nucleon cross section in Higgs portal can provide a spin-independent (SI) cross section, whose sensitivity can be checked with stringent upper bound of LZ-ZEPLIN experiment. The effective interaction Lagrangian in Higgs portal takes the form
\begin{equation}
    \mathcal{L}_{\rm eff} = a_q \phi^R_1 \phi^R_1 q \overline{q}, \quad {\rm where}
\end{equation}
\begin{equation}
    a_q = \frac{M_q}{2 M^2_h M_{R1}} (\lambda_{L1} \cos^2\theta_R + \lambda_{L2}\sin^2\theta_R) ~{\rm with}~ \lambda_{Lj} = \lambda_{Hj}+\lambda^\prime_{Hj}+\lambda^{\prime\prime}_{Hj}.\nonumber
\end{equation}
The corresponding cross section is given by \cite{LopezHonorez:2006gr,Barbieri:2006dq,Dolle:2009fn}
\begin{equation}
    \sigma_{\rm SI} = \frac{1}{4\pi}\left(\frac{M_n M_{R1}}{M_n + M_{R1}}\right)^2 \left(\frac{\lambda_{L1} \cos^2\theta_R + \lambda_{L2}\sin^2\theta_R}{2 M_{R1} M_h^2}\right)^2 f^2 M_n^2,
\end{equation}
where, $M_n$ denotes the nucleon mass, nucleonic matrix element $f \sim 0.3$ \cite{Ellis:2000ds}. We have implemented the model in LanHEP \cite{Semenov:1996es} package and used micrOMEGAs \cite{Pukhov:1999gg, Belanger:2006is, Belanger:2008sj} to compute relic density and also DM-nucleon cross section. The detailed analysis of neutrino and dark matter observables  and their viability through a common parameter space will be discussed in the upcoming section.
\section{Analysis}
\begin{figure}[htb!]
\centering
\includegraphics[scale=0.51]{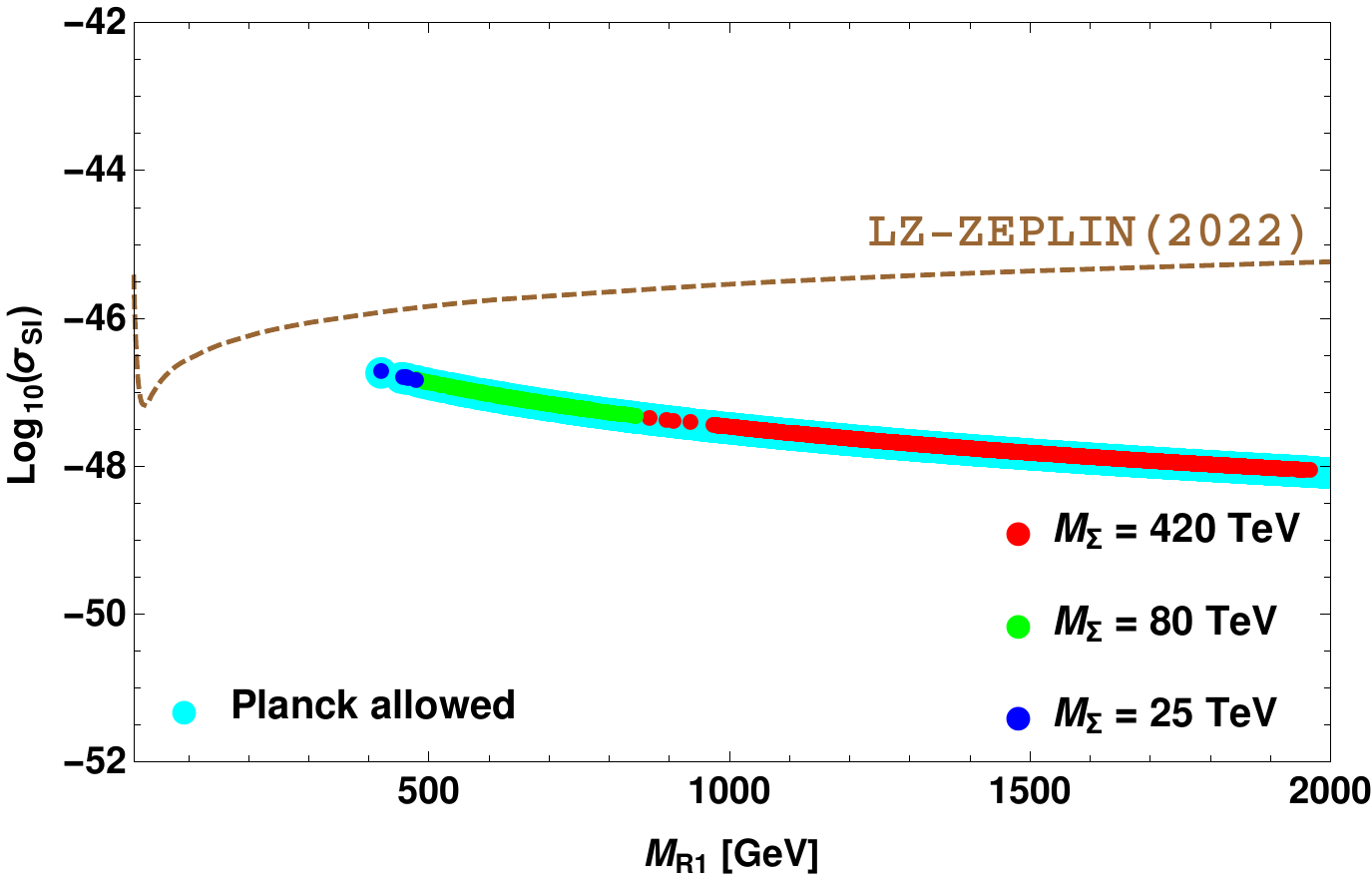}
\hspace{0.5 cm}
\includegraphics[scale=0.57]{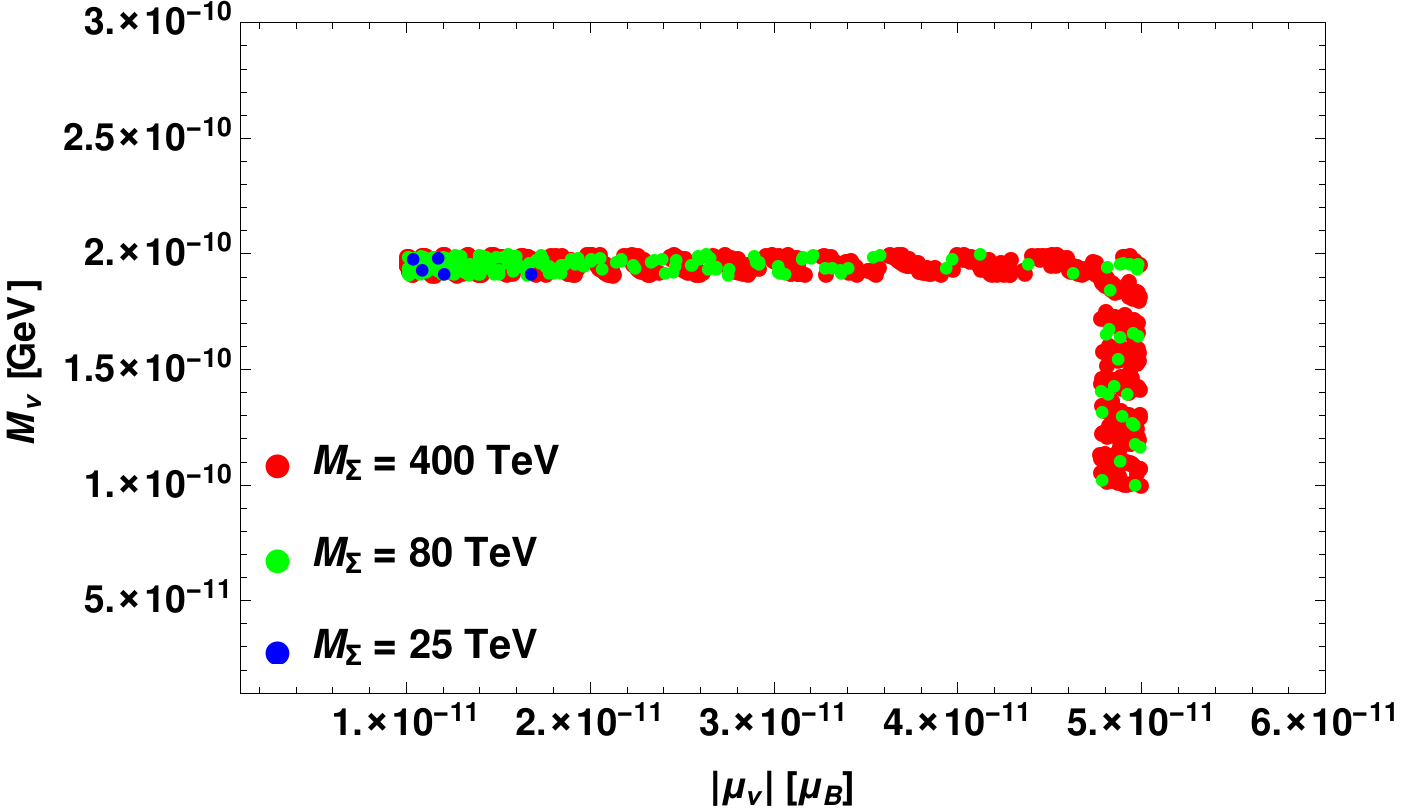}
\caption{Left panel projects SI WIMP-nucleon cross section as a function $M_{R1}$, with dashed brown line of LZ-ZEPLIN upper limit \cite{LZ:2022ufs}. Cyan data points satisfy Planck limit \cite{Aghanim:2018eyx} on relic abundance in $3\sigma$. Blue, green and red data points satisfy neutrino mass and magnetic moment for a specific set of values for fermion triplet and Yukawa, visible in the right panel.}
 \label{scatter1}
\end{figure}

\begin{figure}[htb!]
\centering
\includegraphics[scale=0.45]{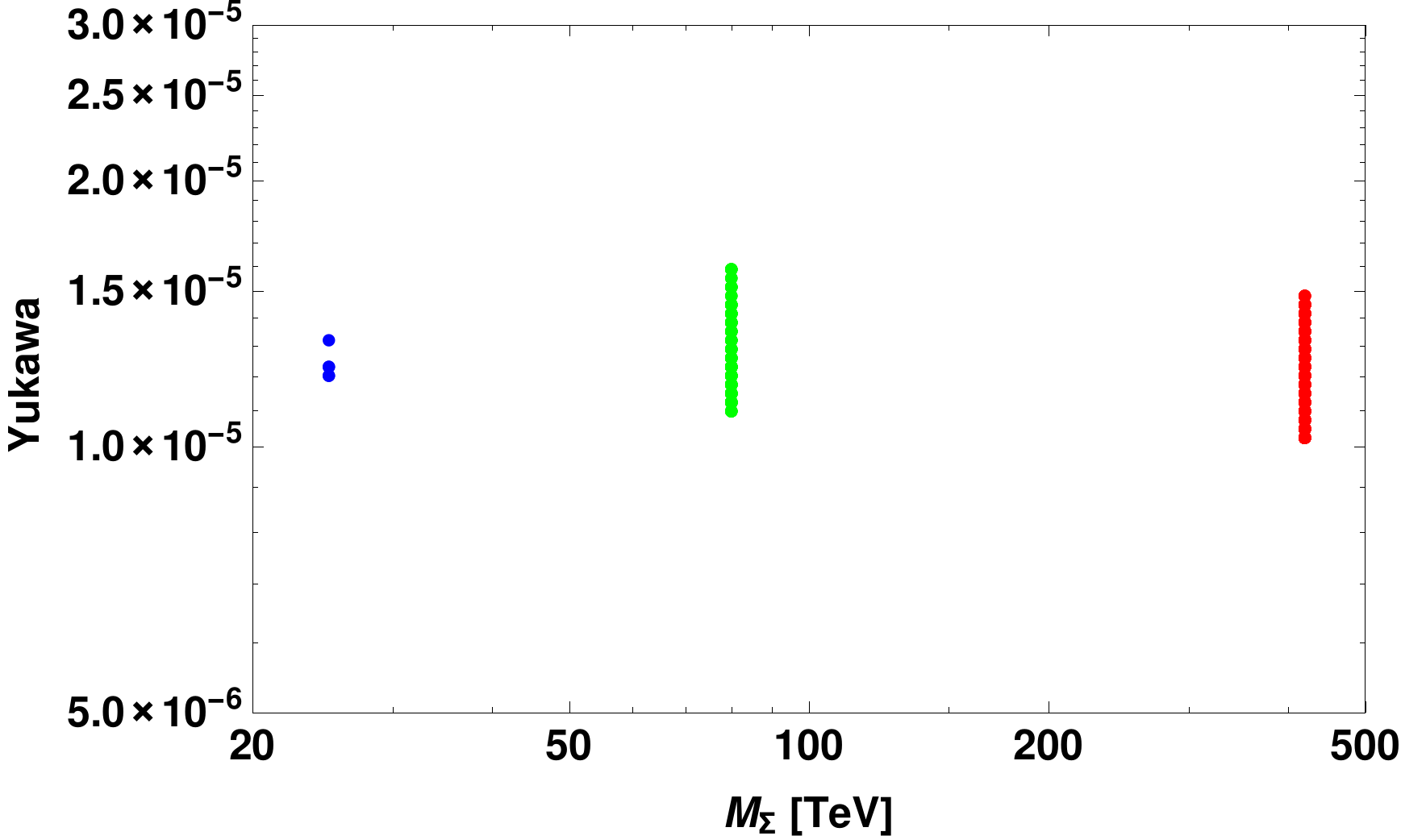}
\hspace{0.5 cm}
\includegraphics[scale=0.4]{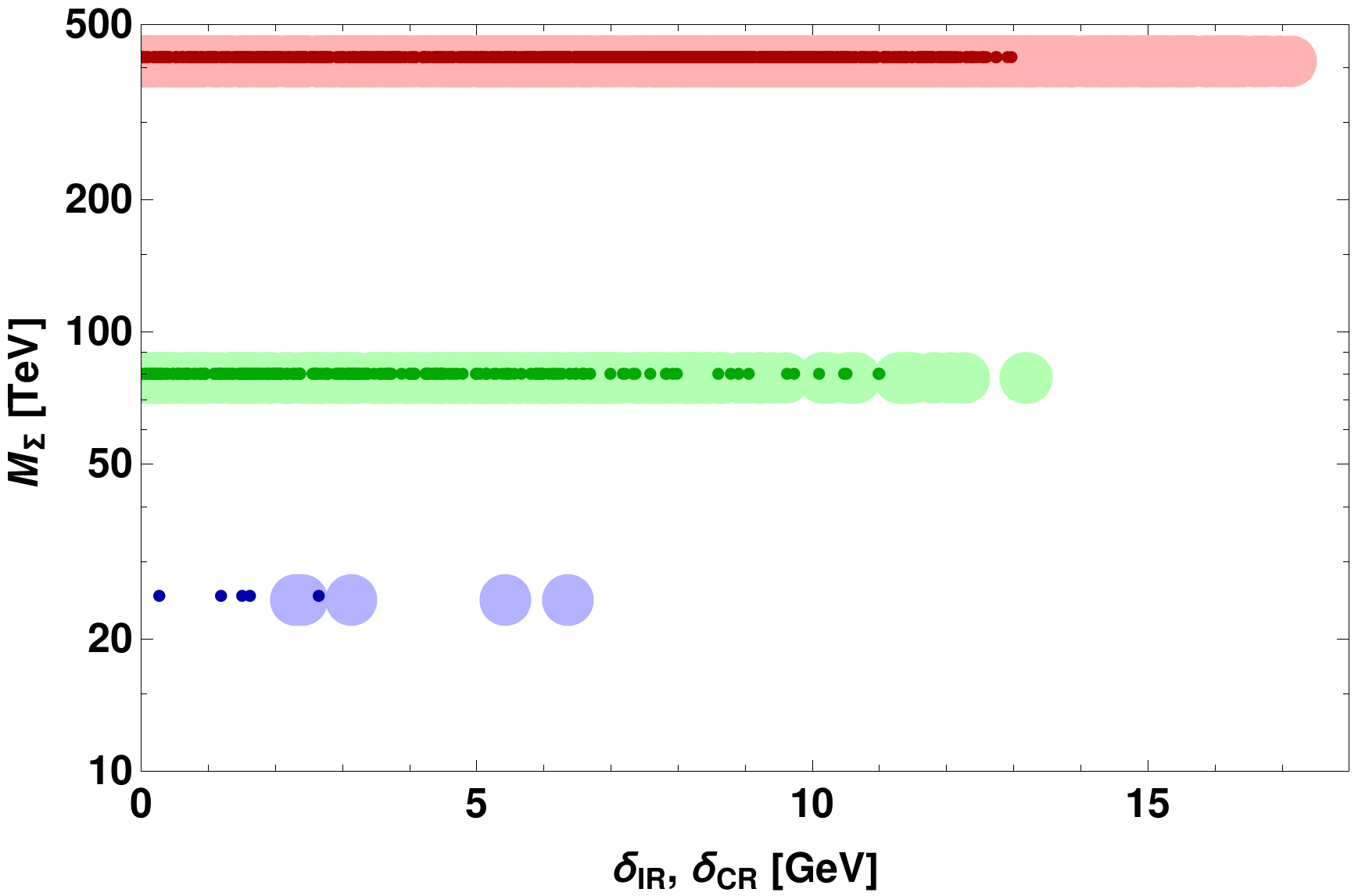}
 \caption{Left panel displays the suitable region for  triplet mass and Yukawa to explain neutrino phenomenology. Right panel shows the allowed region for scalar mass splittings, thick (thin) bands correspond to $\delta_{\rm IR} (\delta_{\rm CR})$ respectively. }
 \label{scatter2}
\end{figure}

\begin{figure}[htb!]
\centering
\includegraphics[scale=0.58]{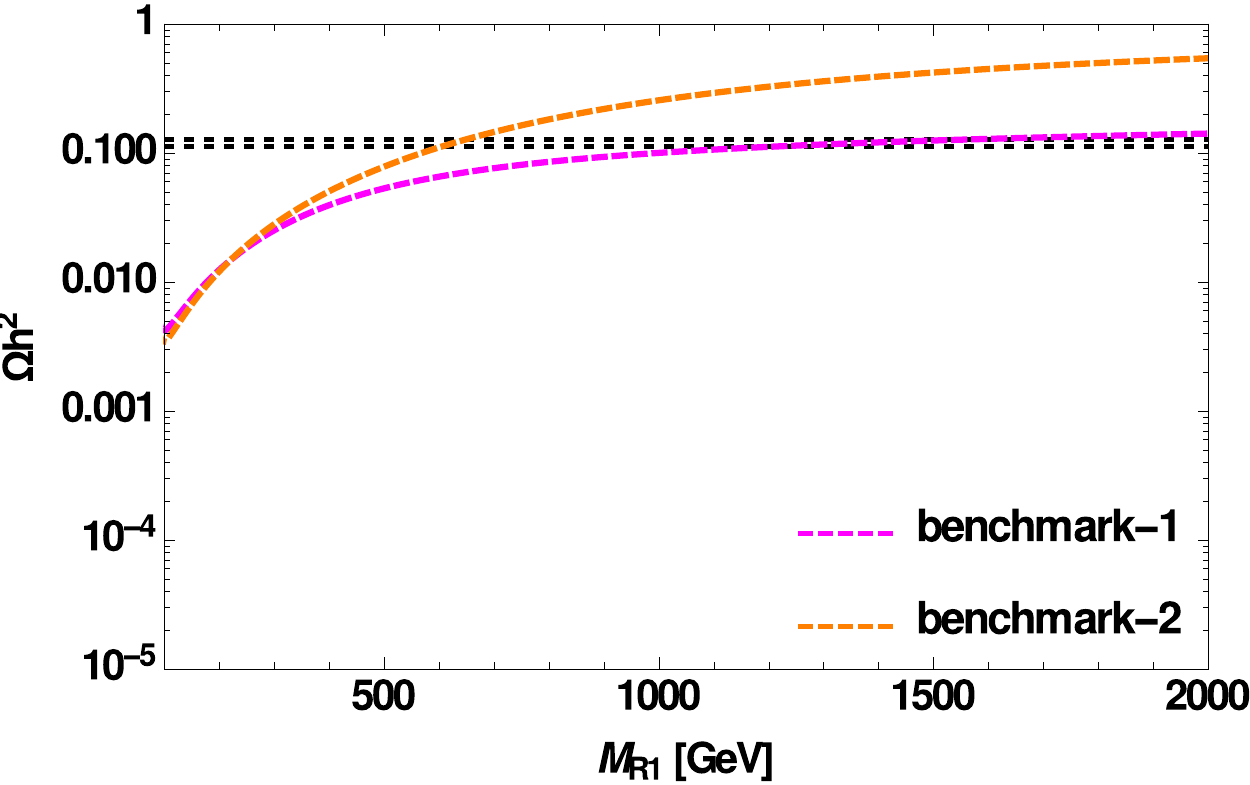}
 \caption{Relic density as a function of dark matter mass for the chosen benchmark of the favourable parameter space of table-\ref{tab_benchmark}.}
 \label{relic_plot}
\end{figure}

In the present framework, we consider $\phi^R_1$ to be the lightest inert scalar eigen state and there are five other heavier scalars. To make the analysis simpler, we consider the mass parameters related to the scalar masses as follows:  one parameter $M_{R1}$ corresponding to the mass of $\phi^R_1$ and three mass splittings namely $\delta$, $\delta_{\rm IR}$ and $\delta_{\rm CR}$. The masses of the rest of the inert scalars can be derived using the following relations:
\begin{eqnarray}
&& M_{R2} - M_{R1} = M_{I2} - M_{I1} = M_{C2} - M_{C1} = \delta, \nn\\
&& M_{Ri} - M_{Ii} = \delta_{\rm IR}, \quad M_{Ri} - M_{Ci} = \delta_{\rm CR}\;,
\end{eqnarray}
where, $i = 1,2$. In the above set up, the scalar mixing angles can be related as follows
\begin{eqnarray}
    \sin2\theta_I = \sin2\theta_R \left(\frac{2 M_{R1} + \delta}{2 M_{R1}+2\delta_{\rm IR} +\delta}\right),\\
    \sin2\theta_C = \sin2\theta_R \left(\frac{2 M_{R1} + \delta}{2 M_{R1}+2\delta_{\rm CR} +\delta}\right).
\end{eqnarray}
We have performed the scan over model parameters as given below, in order to obtain the region, consistent with experimental bounds associated with both dark matter and neutrino sectors: 
\begin{eqnarray}
&&100 ~{\rm GeV} \le M_{R1} \le 2000 ~{\rm GeV}, \quad 0 \le \sin\theta_R \le 1, \nn\\
&&0.1 ~{\rm GeV} \le \delta < 200 ~{\rm GeV}, \quad 0.1 ~{\rm GeV} \le \delta_{\rm IR}, \delta_{\rm CR} \le 20 ~{\rm GeV}.
\end{eqnarray}
We filter out the parameter space by providing Planck constraint on relic density \cite{Aghanim:2018eyx} in $3\sigma$ and then compute DM-nucleon SI cross section for the available parameter space. We project the cross section as a function of $M_{R1}$ in the left panel of Fig. \ref{scatter1} with cyan data points,  where the dashed brown line corresponds to LZ-ZEPLIN upper limit \cite{LZ:2022ufs}. Choosing a set of values for the Yukawa and fermion triplet mass, with the obtained parameter space, one can satisfy the discussed aspects of neutrino phenomenology. The blue, green and red data points corresponding to $25, 80$ and $420$ TeV of triplet mass and suitable Yukawa satisfy the neutrino magnetic moment and light neutrino mass in the desired range simultaneously, as projected in the right panel. We notice that a wide region of dark matter mass is favoured as we move towards high scale (triplet mass) and moreover the favourable region shifts towards larger values with scale. The suitable region of Yukawa and fermion triplet mass is depicted in the left panel of Fig. \ref{scatter2}, allowed range for scalar mass splittings is displayed in the right panel. Here light colored band corresponds to $\delta_{\rm IR}$ and dark colored band stands for $\delta_{\rm CR}$.
Using two specific benchmark values (shown in table. \ref{tab_benchmark} and table. \ref{tab_observables}) which are favourable to explain both neutrino and dark matter aspects discussed so far, we project relic abundance scalar dark matter in Fig. \ref{relic_plot}. 
\begin{table}[htb]
\caption{Set of benchmarks from the consistent parameter space.}
\begin{center}
\begin{tabular}{|c|c|c|c|c|c|c|c|c|c||c|c|}
	\hline
			& $M_{R1}$ [GeV]	& $\delta$ [GeV] & $\delta_{\rm CR}$ [GeV] & $\delta_{\rm IR}$ [GeV] & $M_{\Sigma}$ [TeV]& Yukawa & $\sin\theta_R$ \\
	\hline
	benchmark-1 & $1472$	& $101.69$	& $9.03$ & $0.35$ & $420$ & $10^{-4.89}$ & $0.09$ \\
	\hline
	benchmark-2 & $628$	& $36.40$	& $4.38$ & $3.45$ & $80$ & $10^{-4.85}$ & $0.06$\\
	\hline
\end{tabular}
\label{tab_benchmark}
\end{center}
\end{table}
\begin{table}[htb]
\caption{Neutrino and dark matter observables for the given benchmarks.}
\begin{center}
\begin{tabular}{|c|c|c|c|c|c|c|c|c|c||c|c|}
	\hline
			& $|\mu_\nu|$ [$\mu_B$]& $\mathcal{M}_{\nu}$ [GeV]& ${\rm Log}^{[\sigma_{\rm SI}]}_{10}$ ${\rm cm}^{-2}$ & $\Omega {\rm h}^2$\\
	\hline
	benchmark-1~ & ~$2.73\times 10^{-11}$ & ~$1.99\times 10^{-10}$ & $-47.78$ & $0.123$\\
	\hline
	benchmark-2~ & ~$3.03\times 10^{-11}$ & ~$1.92\times 10^{-10}$ & $-47.04$ & $0.119$\\
	\hline
\end{tabular}
\label{tab_observables}
\end{center}
\end{table}
\vspace*{1.0true cm}

\subsection{Constraints from neutrino oscillation parameters}
More specific constraints on the Yukawa couplings can be obtained from the neutrino oscillation parameters. For this purpose, we consider the neutrino mixing matrix as the product of a tri-bimaximal (TBM) matrix  with a rotation matrix $U_{13}$, given by
\begin{equation}
U_\nu = U_{\rm TBM}\cdot U_{13} = \left(\begin{array}{ccc}
\cos\Theta & \sin\Theta & 0 \\
-\frac{\sin\Theta}{\sqrt 2} & \frac{\cos\Theta}{\sqrt 2} & 
\frac{1}{\sqrt 2}\\
\frac{\sin\Theta}{\sqrt 2} & -\frac{\cos\Theta}{\sqrt 2} & 
\frac{1}{\sqrt 2}\\
\end{array}\right)\cdot   \left ( \begin{array}{ccc}
 \cos \varphi    & 0 & e^{-i  \zeta} \sin \varphi \\
0    & 1    & 0 \\
 -e^{i \zeta} \sin \varphi  & 0   & \cos \varphi\\
\end{array}
\right ).
\end{equation}
From eqns. \ref{mag_eqn},\ref{mass_eqn}, the matrices associated with neutrino magnetic moment and mass can be written in a compact form as
\begin{eqnarray}
    \mu_{\nu} = Y \cdot {\rm diag}(\Lambda_1,\Lambda_2,\Lambda_3)\cdot Y^T,\nn\\
    \mathcal{M}_{\nu} = Y \cdot {\rm diag}(\Lambda^\prime_1,\Lambda^\prime_2,\Lambda^\prime_3) \cdot Y^T,
    \label{eq:28}
\end{eqnarray}
where, 
\begin{eqnarray}
&&\Lambda_k = \frac{1}{8\pi^{2}}M_{\Sigma^+_k}   \bigg[(1+\sin 2\theta_C)\frac{1}{M^{2}_{C2}} \left(\ln \left[\frac{M^{2}_{C2}}{M^2_{\Sigma^+_k}}\right]-1\right)+ (1-\sin 2\theta_C)\frac{1}{M^{2}_{C1}} \left(\ln \left[\frac{M^{2}_{C1}}{M^2_{\Sigma^+_k}}\right]-1\right) \bigg],\nn\\
&&\Lambda^\prime_k = \frac{1}{32 \pi^{2}}M_{\Sigma^+_k}
 \Bigg(\Bigg[(1+\sin2\theta_C)\frac{M_{C2}^2}{M_{\Sigma^+_k}^{2}-M_{C2}^2}\ln\left(\frac{M^2_{\Sigma^+_k}}{M^2_{C2}}\right) + (1-\sin2\theta_C)\frac{M_{C1}^2}{M_{\Sigma^+_k}^{2} -  M_{C1}^2}\ln\left(\frac{M^2_{\Sigma^+_k}}{M^2_{C1}}\right)  \Bigg]  \nonumber\\
&&~~~~~~~~+ \Bigg[(1+\sin2\theta_R)\frac{M_{R2}^2}{M_{\Sigma^0_k}^{2}-M_{R2}^2}\ln\left(\frac{M^2_{\Sigma^0_k}}{M^2_{R2}}\right)+ (1-\sin2\theta_R)\frac{M_{R1}^2}{M_{\Sigma^0_k}^{2} -  M_{R1}^2}\ln\left(\frac{M^2_{\Sigma^0_k}}{M^2_{R1}}\right)  \Bigg]  \nonumber\\
 &&~~~~~~~~-\Bigg[(1+\sin2\theta_I)\frac{M_{I2}^2}{M_{\Sigma^0_k}^{2}-M_{I2}^2}\ln\left(\frac{M^2_{\Sigma^0_k}}{M^2_{I2}}\right)+ (1-\sin2\theta_I)\frac{M_{I1}^2}{M_{\Sigma^0_k}^{2} -  M_{I1}^2}\ln\left(\frac{M^2_{\Sigma^0_k}}{M^2_{I1}}\right)  \Bigg]\Bigg),\nn\\
\end{eqnarray}
and
\begin{equation}
Y = \left(\begin{array}{ccc}
Y_{e1} ~&~ Y_{e2} ~&~ Y_{e3} \\
Y_{\mu 1} ~&~ Y_{\mu 2} ~&~ Y_{\mu 3}\\
Y_{\tau 1} ~&~ Y_{\tau 2} ~&~ Y_{\tau 3}\\
\end{array}\right).
\end{equation}
Diagonalizing the matrices in Eq. (\ref{eq:28}) using $U_{\nu}$, we obtain three unique solutions where the Yukawa couplings corresponding to different flavors become linearly dependent. The relations take the form
\begin{eqnarray}  
&& Y_{e1} = \left(\frac{\sqrt2\;\cos \Theta\,\cos \varphi\,}
{\sin \Theta\,\cos \varphi-e^{-i\zeta}\sin \varphi}\right)Y_{\tau 1}, \hspace{0.1 cm}
Y_{e2} = -\frac{\sqrt2\;\sin \Theta\,}{\cos \Theta} Y_{\tau 2}, \hspace{0.1 cm}
Y_{e3} =
\left(\frac{\sqrt2\;\cos \Theta\,\sin \varphi\,}{\sin \Theta\,\sin \varphi+e^{-i\zeta}\cos \varphi}\right)Y_{\tau 3},
\nonumber \\ 
&& Y_{\mu1} = \left(\frac{e^{-i\zeta}\sin \varphi+\sin \Theta\,\cos \varphi}
{e^{-i\zeta}\sin \varphi-\sin \Theta\,\cos \varphi}\right)Y_{\tau 1}, \hspace{0.2 cm}
Y_{\mu2} = -Y_{\tau 2}, \hspace{0.2 cm}
Y_{\mu3} = \left(\frac{e^{-i\zeta}\cos \varphi-\sin \Theta\,\sin \varphi}
{e^{-i\zeta}\cos \varphi+\sin \Theta\,\sin \varphi}\right)Y_{\tau 3}.
\end{eqnarray}
Thus, the obtained diagonalized matrices associated with neutrino magnetic moment and mass in the basis of active neutrinos are  \cite{Kashiwase:2013uy,Ho:2013hia, Singirala:2016kam}
\begin{equation}
\mu^D_\nu=\left(\begin{array}{ccc}
a_1~ (Y^2_{\tau 1} \Lambda_1) & 0 & 0\\
0 & a_2~ (Y^2_{\tau 2} \Lambda_2) & 0\\
 0 & 0 & a_3 ~(Y^2_{\tau 3} \Lambda_3)\\
\end{array}\right),\quad
\mathcal{M}^D_\nu=\left(\begin{array}{ccc}
a_1~ (Y^2_{\tau 1} \Lambda^\prime_1) & 0 & 0\\
0 & a_2~ (Y^2_{\tau 2} \Lambda^\prime_2) & 0\\
 0 & 0 & a_3 ~(Y^2_{\tau 3} \Lambda^\prime_3)\\
\end{array}\right), 
\label{diagmat}
\end{equation}
where,
\begin{eqnarray}
a_1 = \frac{2 e^{2i\zeta}}{(-e^{i\zeta}\cos \varphi \sin \Theta+\sin \varphi)^2}, \hspace{0.2 cm} a_2 = \frac{2}{\cos^2\Theta}, \hspace{0.2 cm} a_3 = \frac{2 e^{-2i\zeta}}{(e^{-i\zeta}\cos \varphi + \sin \Theta\sin \varphi)^2}.
\end{eqnarray}
The matrix $U_\nu$ replicates the standard Pontecorvo-MakiNakagawa-Sakata (PMNS) matrix, where the mixing angles, $\Theta$ and $\varphi$ can be fixed using the observed neutrino oscillation parameters. Using the best-fit values on $\theta_{12},\theta_{13}$ and $\theta_{23}$ from \cite{Esteban:2020cvm}, we get $\Theta = 33.04^\circ$ and $\varphi = 10.18^\circ$. Furthermore, we take $\zeta = 180^\circ$ which falls within $1\sigma$ region of the observed value of CP phase $197^{+27}_{-24}$ \cite{Esteban:2020cvm}. Substituting the above, we get, 
\begin{equation}
|U_\nu|=\left(\begin{array}{ccc}
0.8250 ~&~ 0.5452 ~&~ 0.1481 \\
0.2544 ~&~ 0.5927 ~&~ 0.7641\\
0.5044 ~&~ 0.5927 ~&~ 0.6278\\
\end{array}\right),
\end{equation}
which is consistent enough in comparison with the leptonic mixing matrix that can explain the observed mixing angles in $3\sigma$ region \cite{Esteban:2020cvm} 
\begin{equation}
|U^{{\rm SK}-3\sigma}_{\rm PMNS}|=\left(\begin{array}{ccc}
0.801 \to 0.845 ~&~ 0.513 \to 0.579 ~&~ 0.143 \to 0.155 \\
0.234 \to 0.500 ~&~ 0.471 \to 0.689 ~&~ 0.637 \to 0.776\\
0.271 \to 0.525 ~&~ 0.477 \to 0.694 ~&~ 0.613 \to 0.756\\
\end{array}\right).
\end{equation}
Furthermore, the Yukawa matrix turns out to be
\begin{equation}
Y=\left(\begin{array}{ccc}
1.6356~ Y_{\tau 1} & -0.9197~ Y_{\tau 2} & -0.2360~ Y_{\tau 3}\\
-0.5044~ Y_{\tau 1}& -Y_{\tau 2} & 1.2170~ Y_{\tau 3}\\
 Y_{\tau 1} & Y_{\tau 2} & Y_{\tau 3}\\
\end{array}\right),
\end{equation}
and the relevant coefficients in the diagonal matrices of eqn. \ref{diagmat} become $a_1 = 3.930$, $a_2 = 2.846$ and $a_3 = 2.537$.

\begin{figure}[htb!]
\centering
\includegraphics[scale=0.6]{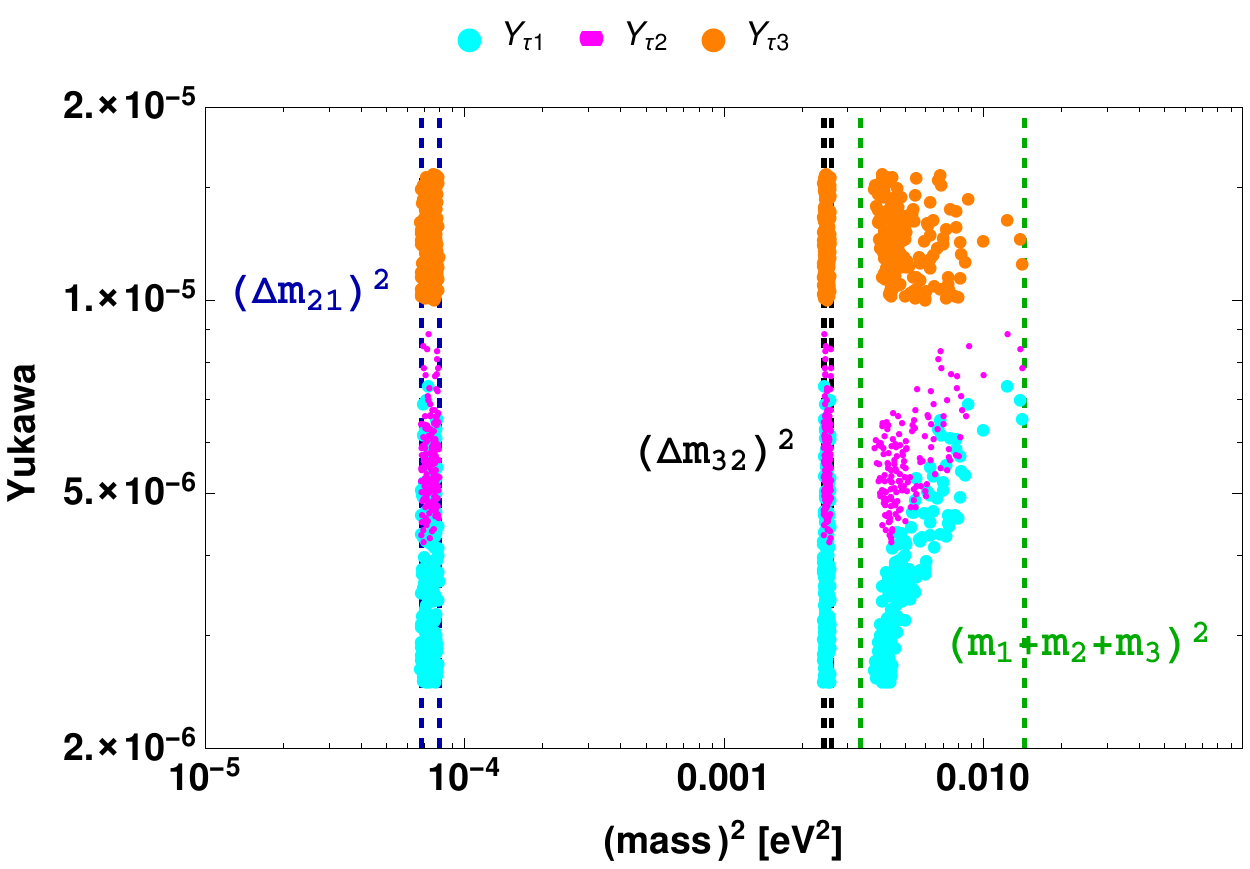}
 \caption{Allowed region of Yukawa (colored data points), satisfying the $3\sigma$ limit on mass squared differences \cite{Esteban:2020cvm} and cosmological bounds on sum of active neutrino masses \cite{RoyChoudhury:2018gay} (colored vertical lines).} 
 \label{yuk_osc}
\end{figure}

To illustrate, we consider all the three generations of heavy fermion triplets to be degenerate in mass i.e., $m_\Sigma \sim 420$ TeV and scan the DM consistent parameter space to extract the constraints on Yukawa. In Fig. \ref{yuk_osc}, we project the allowed region of Yukawa (shown as cyan, magenta and orange data points) that satisfies the required bound on mass squared differences \cite{Esteban:2020cvm} in $3\sigma$ (blue and black dashed lines) and also cosmological bound on sum of active neutrino masses i.e., $0.058 \le (m_1+m_2+m_3) \le 0.12$ eV (normal hierarchy) \cite{RoyChoudhury:2018gay} represented as green vertical lines. To mention, this projected parameter space is consistent with neutrino magnetic moment in a wide range of values, i.e., $3\times 10^{-12}$ to $7 \times 10^{-10}$ (in units of $\mu_B$), which is the regime where all the experiments have provided the bounds (to be discussed in the next section).
\section{Implications}
In the experimental perspective, non-zero neutrino magnetic moment of solar neutrinos can provide explanation for the excess in electron recoil events at XENON1T collaboration \cite{XENON:2020rca}. In other words, the neutrino transition magnetic moment can provide additional contribution to the neutrino-electron scattering process. In this section, we utilize non-zero transition neutrino magnetic moment to explain the excess in electron recoil events. 
 
\begin{figure}[htb!]
\centering
\includegraphics[scale=0.6]{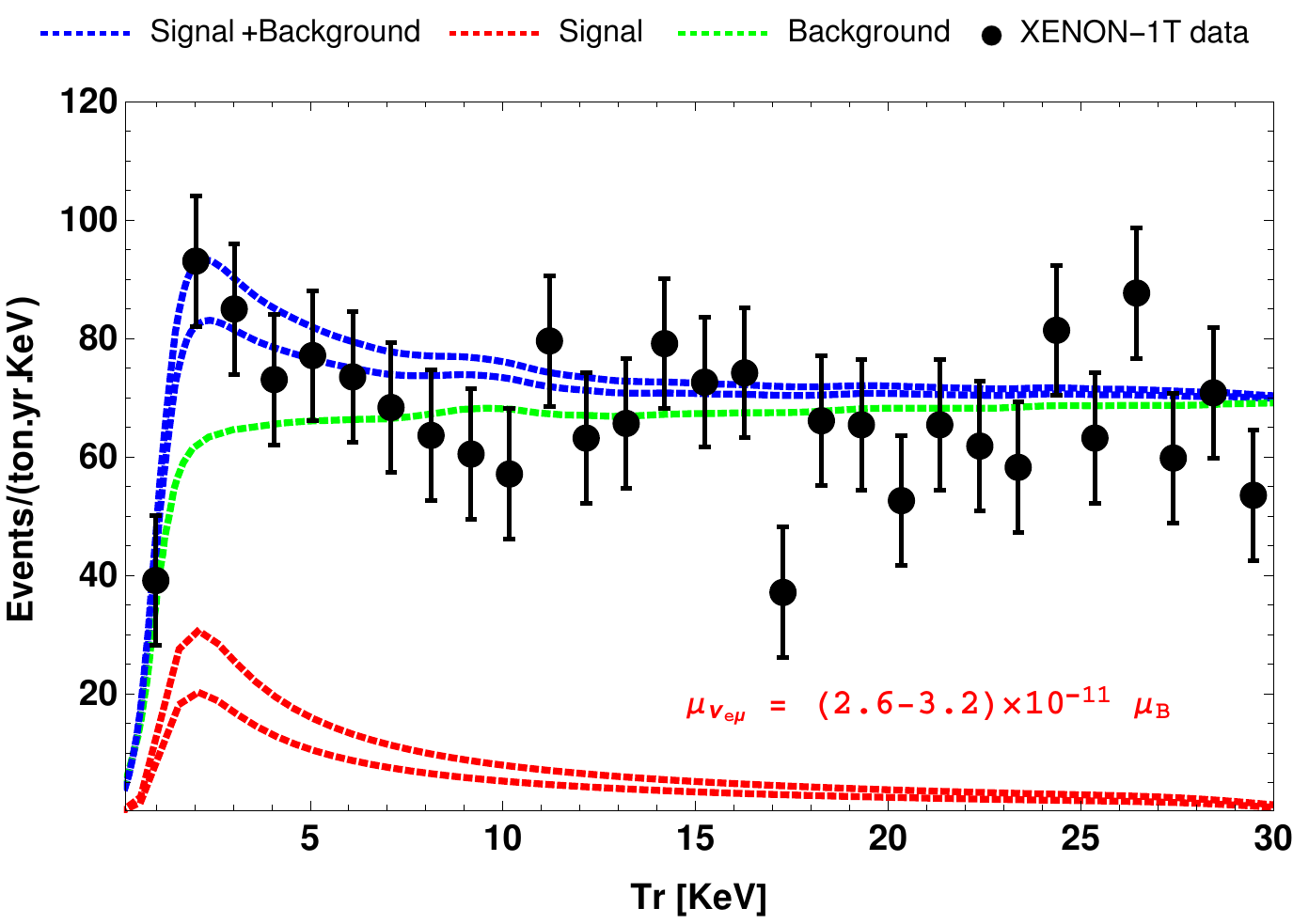}
 \caption{Excess recoil events (blue) to match XENON1T \cite{XENON:2020rca} (black) through the signal from transition neutrino magnetic moment $\mu_{\nu_{e\mu}}= 2.6 \times 10^{-11} \mu_B$ and $3.2 \times 10^{-11} \mu_B$ (in red) along with background (green).}
 \label{event excess}
\end{figure}

In the presence of magnetic moment, the total differential cross section can be written as \cite{Giunti:2015gga}
\begin{eqnarray}
    \left( \frac{d \sigma}{d T_{e}} \right)_{\rm TOT} =  \left( \frac{d \sigma}{d T_{e}} \right)_{\rm SM} + \left( \frac{d \sigma}{d T_{e}} \right)_{\rm EM},
    \label{cross-section}
\end{eqnarray}
where $T_{e}$ is the electron recoil energy. The first contribution in eq. \ref{cross-section} is due to standard weak interactions, given by
\begin{equation}
\left( \frac{d \sigma}{d T_{e}} \right)_{\rm SM} =  \frac{G_F^2 m_e}{2\pi} \left[(g_V + g_A)^2 + \left(1-\frac{T_e}{E_\nu}\right)^2(g_V - g_A)^2 + \left(\frac{m_e T_e}{E_\nu^2}\right)(g^2_A - g^2_V)\right].
\end{equation}
In the above, $G_F$ stands for the Fermi constant and
\begin{eqnarray}
&&g_V = 2\sin^2 \theta_W + \frac{1}{2},~~ g_A = 1/2 ~~{\rm for}~~ \nu_e, \nn\\
&&g_V = 2\sin^2 \theta_W - \frac{1}{2},~~ g_A = -1/2 ~~{\rm for}~~ \nu_\mu, \nu_\tau.
\end{eqnarray}
The second contribution comes from the effective electromagnetic vertex of the neutrinos, i.e.,  magnetic moment contribution, which is expressed as
\begin{eqnarray}
    \left( \frac{d \sigma}{d T_{e}} \right)_{\rm EM} = \frac{\pi \alpha^{2}}{m^{2}_{e}} \left( \frac{1}{T_{e}} - \frac{1}{E_{\nu}} \right) \left( \frac{\mu_{\nu_{e\mu}}}{\mu_{B}} \right)^{2},
\end{eqnarray}
where, $\alpha$ is the electromagnetic coupling, $E_{\nu}$ is the initial neutrino energy, $\mu_{\nu_{e \mu}}$ is the neutrino magnetic moment and $\mu_{B}$ is the Bohr magneton. For high $T_{e}$ value, weak cross-section dominates and for low $T_{e}$ value, the electromagnetic cross-section dominates and hence, we search for the signature of neutrino magnetic moment in the low energy region. For simplicity, we take one transition magnetic moment $\mu_{{\nu}_{e \mu}}$ to explain the XENON1T excess. The differential event rate to estimate the XENON1T signal is given by
\begin{equation}
\frac{dN}{ dT_r} =  n_{\rm te}\times \int^{E^{\rm max}_\nu}_{E^{\rm min}_\nu} dE_\nu \int_{T_{\rm th}}^{T_{\rm max}} dT_e \left(\frac{d\sigma^{\nu_e e}}{dT_e} P_{ee} + \cos^2 \theta_{23} \frac{d\sigma^{\nu_\mu e}}{dT_e} P_{e\mu}\right) \times \frac{d\phi_s}{dE_\nu} \times \epsilon(T_e)\times G(T_e,T_r).
\end{equation}
In the above, $\epsilon(T_e)$ denotes the efficiency of detector \cite{XENON:2020rca}, $n_{te}$ is the count of number of target electrons in the fiducial volume of one ton Xenon \cite{XENON:2020iwh}, $d\phi_s/dE_\nu$ represents the solar neutrino flux spectrum \cite{Bahcall:2004mz}, and the function $G(T_e,T_r)$ reflects the normalised Gaussian smearing function that takes into account the detector's limited energy resolution \cite{XENON:2020rca}. 
The limits $T_{\rm th} = 1$ KeV and $T_{\rm max} =  30$ KeV stand for the threshold and maximum recoil energy of detector respectively. The extremes of neutrino energy for the integral are given by $E^{\rm max}_\nu = 420$ KeV and $E^{\rm min}_\nu = [T+(2m_e T+T^2)^\frac{1}{2}]/2$ \cite{Babu:2020ivd}. The survival probability can be expressed as 
\begin{equation}
    P_{ee} = \sin^4 \theta_{13} + \frac{1}{2} \cos^4 \theta_{13} (1 + \cos 2 \theta^{m}_{12} + \cos 2 \theta_{12})\;.
\end{equation}
And the disappearance probability can be taken as $P_{e \mu /\tau}$ = $1 - P_{ee}$ \cite{Lopes:2013nfa, ParticleDataGroup:2020ssz}. The oscillation parameters are taken from \cite{Esteban:2020cvm}. In Fig. \ref{event excess}, we project the event rate as a function of recoil energy $T_r$, for two set of values for magnetic moment, i.e., $\mu_{\nu_{e\mu}}=2.6\times 10^{-11}\mu_B $  and  $3.2\times 10^{-11}\mu_B $ (red curves). Adding with the background (green curve), we are able to meet the observed recoil event excess in the low energy region near $2.5$ KeV as of XENON1T experiment \cite{XENON:2020rca}.

\begin{figure}[htb!]
\centering
\includegraphics[scale=0.6]{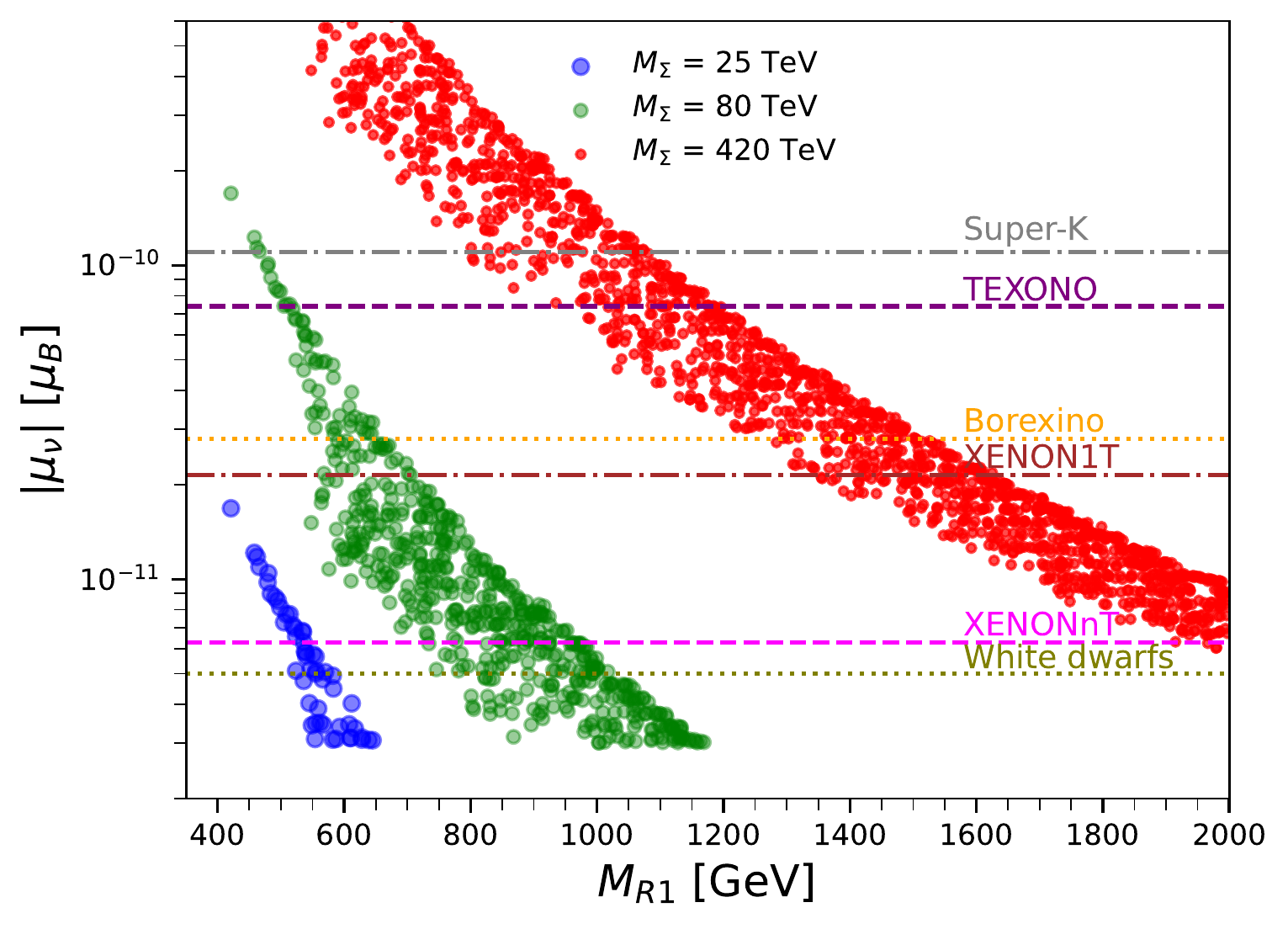}
 \caption{Allowed region of neutrino magnetic moment with the  mass spectrum of dark matter and fermion triplet. Horizontal colored lines stand for the upper bounds of XENON1T (average value of suggested range in \cite{XENON:2020rca}), XENONnT \cite{XENON:2022ltv}, Borexino \cite{Borexino:2017fbd}, TEXONO \cite{TEXONO:2006xds}, Super-K \cite{Super-Kamiokande:2004wqk} and white dwarfs \cite{MillerBertolami:2014oki}.}
 \label{allexp_plot}
\end{figure}

In Fig. \ref{allexp_plot}, we project neutrino magnetic moment as a function of dark matter mass, choosing specific set of values assigned to triplet fermion. As seen earlier in the left panel of Fig. \ref{scatter1}, a specific range of dark matter mass is favoured with the scale of triplet mass. It is transparent that the model parameters are able to provide neutrino magnetic moment in the range $10^{-12}\mu_B$ to $10^{-10}\mu_B$, sensitive to the upper limits of Super-K \cite{Super-Kamiokande:2004wqk}, TEXONO \cite{TEXONO:2006xds}, Borexino \cite{Borexino:2017fbd}, XENON1T \cite{XENON:2020rca}, XENONnT \cite{XENON:2022ltv} and white dwarfs \cite{MillerBertolami:2014oki} (colored horizontal lines). Thus, from all the above discussions made, it is evident that this simple framework can provide a consistent phenomenological platform for a correlative study of neutrino magnetic moment (especially), mass and dark matter physics.
\section{Concluding remarks}
The primary aim is to address neutrino mass, magnetic moment and dark matter phenomenology in a common framework. For this purpose, we have extended the standard model with three vector-like fermion triplets and two inert scalar doublets to realize Type-III radiative scenario.  A pair of charged scalars help in obtaining neutrino magnetic moment, all charged and neutral scalars come up in getting light neutrino mass at one-loop level. All the inert scalars participate in annihilation and co-annihilation channels to provide total dark matter relic density of the Universe, consistent with Planck satellite data and also provide a suitable cross section with nucleon, sensitive to LZ-ZEPLIN upper limit. The lightest dark matter mass is scanned upto $2$ TeV while the fermion triplet mass is taken with larger values, i.e., few hundred TeV. Choosing Yukawa of the order $10^{-5}$, we have obtained light neutrino mass in sub-eV scale and also transition magnetic moment  $\sim 10^{-11}$ $\mu_B$, to  successfully explain the excess electron recoil events at low energy scale reported by XENON1T experiment. Finally, we have also demonstrated with a plot that the model is able to provide neutrino magnetic moment in a wide range ($10^{-12}\mu_B$ to $10^{-10}\mu_B$), in the same ball park of Borexino, Super-K, TEXONO, XENONnT and white dwarfs. Overall, this simple model provides a suitable platform to study neutrino phenomenology, especially the neutrino magnetic moment and also dark matter aspects.

\acknowledgments 
SS and RM would like to acknowledge University of Hyderabad IoE project grant no. RC1-20-012.  DKS would like to acknowledge Prime Minister's Research Fellowship, Govt. of India. SS would like to thank Dr. Siddhartha Karmakar and Papia Panda for helpful input in the work.


\bibliography{nmm_DM}

\begin{thebibliography}{65}
\expandafter\ifx\csname natexlab\endcsname\relax\def\natexlab#1{#1}\fi
\expandafter\ifx\csname bibnamefont\endcsname\relax
  \def\bibnamefont#1{#1}\fi
\expandafter\ifx\csname bibfnamefont\endcsname\relax
  \def\bibfnamefont#1{#1}\fi
\expandafter\ifx\csname citenamefont\endcsname\relax
  \def\citenamefont#1{#1}\fi
\expandafter\ifx\csname url\endcsname\relax
  \def\url#1{\texttt{#1}}\fi
\expandafter\ifx\csname urlprefix\endcsname\relax\def\urlprefix{URL }\fi
\providecommand{\bibinfo}[2]{#2}
\providecommand{\eprint}[2][]{\url{#2}}

\bibitem[{\citenamefont{Bilenky}(1999)}]{Bilenky:1999ge}
\bibinfo{author}{\bibfnamefont{S.~M.} \bibnamefont{Bilenky}}, in
  \emph{\bibinfo{booktitle}{{1999 European School of High-Energy Physics}}}
  (\bibinfo{year}{1999}), pp. \bibinfo{pages}{187--217},
  \eprint{hep-ph/0001311}.

\bibitem[{\citenamefont{Mohapatra and Senjanovic}(1980)}]{Mohapatra:1979ia}
\bibinfo{author}{\bibfnamefont{R.~N.} \bibnamefont{Mohapatra}}
  \bibnamefont{and}
  \bibinfo{author}{\bibfnamefont{G.}~\bibnamefont{Senjanovic}},
  \bibinfo{journal}{Phys. Rev. Lett.} \textbf{\bibinfo{volume}{44}},
  \bibinfo{pages}{912} (\bibinfo{year}{1980}).

\bibitem[{\citenamefont{Schechter and Valle}(1980)}]{Schechter:1980gr}
\bibinfo{author}{\bibfnamefont{J.}~\bibnamefont{Schechter}} \bibnamefont{and}
  \bibinfo{author}{\bibfnamefont{J.~W.~F.} \bibnamefont{Valle}},
  \bibinfo{journal}{Phys. Rev. D} \textbf{\bibinfo{volume}{22}},
  \bibinfo{pages}{2227} (\bibinfo{year}{1980}).

\bibitem[{\citenamefont{Babu and Mohapatra}(1993)}]{Babu:1992ia}
\bibinfo{author}{\bibfnamefont{K.}~\bibnamefont{Babu}} \bibnamefont{and}
  \bibinfo{author}{\bibfnamefont{R.}~\bibnamefont{Mohapatra}},
  \bibinfo{journal}{Phys. Rev. Lett.} \textbf{\bibinfo{volume}{70}},
  \bibinfo{pages}{2845} (\bibinfo{year}{1993}), \eprint{hep-ph/9209215}.

\bibitem[{\citenamefont{Hosaka et~al.}(2006)}]{Super-Kamiokande:2005wtt}
\bibinfo{author}{\bibfnamefont{J.}~\bibnamefont{Hosaka}} \bibnamefont{et~al.}
  (\bibinfo{collaboration}{Super-Kamiokande}), \bibinfo{journal}{Phys. Rev. D}
  \textbf{\bibinfo{volume}{73}}, \bibinfo{pages}{112001}
  (\bibinfo{year}{2006}), \eprint{hep-ex/0508053}.

\bibitem[{\citenamefont{Ahmad et~al.}(2002)}]{SNO:2002tuh}
\bibinfo{author}{\bibfnamefont{Q.~R.} \bibnamefont{Ahmad}} \bibnamefont{et~al.}
  (\bibinfo{collaboration}{SNO}), \bibinfo{journal}{Phys. Rev. Lett.}
  \textbf{\bibinfo{volume}{89}}, \bibinfo{pages}{011301}
  (\bibinfo{year}{2002}), \eprint{nucl-ex/0204008}.

\bibitem[{\citenamefont{Abe et~al.}(2016)}]{Super-Kamiokande:2016yck}
\bibinfo{author}{\bibfnamefont{K.}~\bibnamefont{Abe}} \bibnamefont{et~al.}
  (\bibinfo{collaboration}{Super-Kamiokande}), \bibinfo{journal}{Phys. Rev. D}
  \textbf{\bibinfo{volume}{94}}, \bibinfo{pages}{052010}
  (\bibinfo{year}{2016}), \eprint{1606.07538}.

\bibitem[{\citenamefont{Abe et~al.}(2019)}]{T2K:2019efw}
\bibinfo{author}{\bibfnamefont{K.}~\bibnamefont{Abe}} \bibnamefont{et~al.}
  (\bibinfo{collaboration}{T2K}), \bibinfo{journal}{Phys. Rev. D}
  \textbf{\bibinfo{volume}{99}}, \bibinfo{pages}{071103}
  (\bibinfo{year}{2019}), \eprint{1902.06529}.

\bibitem[{\citenamefont{An et~al.}(2012)}]{DayaBay:2012fng}
\bibinfo{author}{\bibfnamefont{F.~P.} \bibnamefont{An}} \bibnamefont{et~al.}
  (\bibinfo{collaboration}{Daya Bay}), \bibinfo{journal}{Phys. Rev. Lett.}
  \textbf{\bibinfo{volume}{108}}, \bibinfo{pages}{171803}
  (\bibinfo{year}{2012}), \eprint{1203.1669}.

\bibitem[{\citenamefont{Abe et~al.}(2012)}]{DoubleChooz:2011ymz}
\bibinfo{author}{\bibfnamefont{Y.}~\bibnamefont{Abe}} \bibnamefont{et~al.}
  (\bibinfo{collaboration}{Double Chooz}), \bibinfo{journal}{Phys. Rev. Lett.}
  \textbf{\bibinfo{volume}{108}}, \bibinfo{pages}{131801}
  (\bibinfo{year}{2012}), \eprint{1112.6353}.

\bibitem[{\citenamefont{Zwicky}(1937)}]{Zwicky:1937zza}
\bibinfo{author}{\bibfnamefont{F.}~\bibnamefont{Zwicky}},
  \bibinfo{journal}{Astrophys. J.} \textbf{\bibinfo{volume}{86}},
  \bibinfo{pages}{217} (\bibinfo{year}{1937}).

\bibitem[{\citenamefont{Zwicky}(1933)}]{PhysRev.43.147}
\bibinfo{author}{\bibfnamefont{F.}~\bibnamefont{Zwicky}},
  \bibinfo{journal}{Phys. Rev.} \textbf{\bibinfo{volume}{43}},
  \bibinfo{pages}{147} (\bibinfo{year}{1933}),
  \urlprefix\url{https://link.aps.org/doi/10.1103/PhysRev.43.147}.

\bibitem[{\citenamefont{Bertone et~al.}(2005)\citenamefont{Bertone, Hooper, and
  Silk}}]{Bertone:2004pz}
\bibinfo{author}{\bibfnamefont{G.}~\bibnamefont{Bertone}},
  \bibinfo{author}{\bibfnamefont{D.}~\bibnamefont{Hooper}}, \bibnamefont{and}
  \bibinfo{author}{\bibfnamefont{J.}~\bibnamefont{Silk}},
  \bibinfo{journal}{Phys. Rept.} \textbf{\bibinfo{volume}{405}},
  \bibinfo{pages}{279} (\bibinfo{year}{2005}), \eprint{hep-ph/0404175}.

\bibitem[{\citenamefont{Mambrini et~al.}(2015)\citenamefont{Mambrini, Nagata,
  Olive, Quevillon, and Zheng}}]{Mambrini:2015vna}
\bibinfo{author}{\bibfnamefont{Y.}~\bibnamefont{Mambrini}},
  \bibinfo{author}{\bibfnamefont{N.}~\bibnamefont{Nagata}},
  \bibinfo{author}{\bibfnamefont{K.~A.} \bibnamefont{Olive}},
  \bibinfo{author}{\bibfnamefont{J.}~\bibnamefont{Quevillon}},
  \bibnamefont{and} \bibinfo{author}{\bibfnamefont{J.}~\bibnamefont{Zheng}},
  \bibinfo{journal}{Phys. Rev. D} \textbf{\bibinfo{volume}{91}},
  \bibinfo{pages}{095010} (\bibinfo{year}{2015}), \eprint{1502.06929}.

\bibitem[{\citenamefont{Sakharov}(1991)}]{Sakharov:1967dj}
\bibinfo{author}{\bibfnamefont{A.}~\bibnamefont{Sakharov}},
  \bibinfo{journal}{Sov. Phys. Usp.} \textbf{\bibinfo{volume}{34}},
  \bibinfo{pages}{392} (\bibinfo{year}{1991}).

\bibitem[{\citenamefont{Kolb and Wolfram}(1980)}]{Kolb:1979qa}
\bibinfo{author}{\bibfnamefont{E.~W.} \bibnamefont{Kolb}} \bibnamefont{and}
  \bibinfo{author}{\bibfnamefont{S.}~\bibnamefont{Wolfram}},
  \bibinfo{journal}{Nucl. Phys. B} \textbf{\bibinfo{volume}{172}},
  \bibinfo{pages}{224} (\bibinfo{year}{1980}), \bibinfo{note}{[Erratum:
  Nucl.Phys.B 195, 542 (1982)]}.

\bibitem[{\citenamefont{Fukugita and Yanagida}(1986)}]{Fukugita:1986hr}
\bibinfo{author}{\bibfnamefont{M.}~\bibnamefont{Fukugita}} \bibnamefont{and}
  \bibinfo{author}{\bibfnamefont{T.}~\bibnamefont{Yanagida}},
  \bibinfo{journal}{Phys. Lett. B} \textbf{\bibinfo{volume}{174}},
  \bibinfo{pages}{45} (\bibinfo{year}{1986}).

\bibitem[{\citenamefont{Fritzsch and Minkowski}(1975)}]{Fritzsch:1974nn}
\bibinfo{author}{\bibfnamefont{H.}~\bibnamefont{Fritzsch}} \bibnamefont{and}
  \bibinfo{author}{\bibfnamefont{P.}~\bibnamefont{Minkowski}},
  \bibinfo{journal}{Annals Phys.} \textbf{\bibinfo{volume}{93}},
  \bibinfo{pages}{193} (\bibinfo{year}{1975}).

\bibitem[{\citenamefont{Bifani et~al.}(2019)\citenamefont{Bifani,
  Descotes-Genon, Romero~Vidal, and Schune}}]{Bifani:2018zmi}
\bibinfo{author}{\bibfnamefont{S.}~\bibnamefont{Bifani}},
  \bibinfo{author}{\bibfnamefont{S.}~\bibnamefont{Descotes-Genon}},
  \bibinfo{author}{\bibfnamefont{A.}~\bibnamefont{Romero~Vidal}},
  \bibnamefont{and} \bibinfo{author}{\bibfnamefont{M.-H.}
  \bibnamefont{Schune}}, \bibinfo{journal}{J. Phys. G}
  \textbf{\bibinfo{volume}{46}}, \bibinfo{pages}{023001}
  (\bibinfo{year}{2019}), \eprint{1809.06229}.

\bibitem[{\citenamefont{Aprile et~al.}(2020{\natexlab{a}})}]{XENON:2020rca}
\bibinfo{author}{\bibfnamefont{E.}~\bibnamefont{Aprile}} \bibnamefont{et~al.}
  (\bibinfo{collaboration}{XENON}), \bibinfo{journal}{Phys. Rev. D}
  \textbf{\bibinfo{volume}{102}}, \bibinfo{pages}{072004}
  (\bibinfo{year}{2020}{\natexlab{a}}), \eprint{2006.09721}.

\bibitem[{\citenamefont{Aprile et~al.}(2022)}]{XENON:2022ltv}
\bibinfo{author}{\bibfnamefont{E.}~\bibnamefont{Aprile}} \bibnamefont{et~al.}
  (\bibinfo{collaboration}{XENON}), \bibinfo{journal}{Phys. Rev. Lett.}
  \textbf{\bibinfo{volume}{129}}, \bibinfo{pages}{161805}
  (\bibinfo{year}{2022}), \eprint{2207.11330}.

\bibitem[{\citenamefont{Miranda et~al.}(2020)\citenamefont{Miranda, Papoulias,
  T\'ortola, and Valle}}]{Miranda:2020kwy}
\bibinfo{author}{\bibfnamefont{O.~G.} \bibnamefont{Miranda}},
  \bibinfo{author}{\bibfnamefont{D.~K.} \bibnamefont{Papoulias}},
  \bibinfo{author}{\bibfnamefont{M.}~\bibnamefont{T\'ortola}},
  \bibnamefont{and} \bibinfo{author}{\bibfnamefont{J.~W.~F.}
  \bibnamefont{Valle}}, \bibinfo{journal}{Phys. Lett. B}
  \textbf{\bibinfo{volume}{808}}, \bibinfo{pages}{135685}
  (\bibinfo{year}{2020}), \eprint{2007.01765}.

\bibitem[{\citenamefont{Li and Xia}(2022)}]{Li:2022bqr}
\bibinfo{author}{\bibfnamefont{Y.-F.} \bibnamefont{Li}} \bibnamefont{and}
  \bibinfo{author}{\bibfnamefont{S.-y.} \bibnamefont{Xia}},
  \bibinfo{journal}{Phys. Rev. D} \textbf{\bibinfo{volume}{106}},
  \bibinfo{pages}{095022} (\bibinfo{year}{2022}), \eprint{2203.16525}.

\bibitem[{\citenamefont{Miranda et~al.}(2021)\citenamefont{Miranda, Papoulias,
  Sanders, T\'ortola, and Valle}}]{Miranda:2021kre}
\bibinfo{author}{\bibfnamefont{O.~G.} \bibnamefont{Miranda}},
  \bibinfo{author}{\bibfnamefont{D.~K.} \bibnamefont{Papoulias}},
  \bibinfo{author}{\bibfnamefont{O.}~\bibnamefont{Sanders}},
  \bibinfo{author}{\bibfnamefont{M.}~\bibnamefont{T\'ortola}},
  \bibnamefont{and} \bibinfo{author}{\bibfnamefont{J.~W.~F.}
  \bibnamefont{Valle}}, \bibinfo{journal}{JHEP} \textbf{\bibinfo{volume}{12}},
  \bibinfo{pages}{191} (\bibinfo{year}{2021}), \eprint{2109.09545}.

\bibitem[{\citenamefont{Babu et~al.}(2021)\citenamefont{Babu, Jana, Lindner,
  and K}}]{Babu:2021jnu}
\bibinfo{author}{\bibfnamefont{K.~S.} \bibnamefont{Babu}},
  \bibinfo{author}{\bibfnamefont{S.}~\bibnamefont{Jana}},
  \bibinfo{author}{\bibfnamefont{M.}~\bibnamefont{Lindner}}, \bibnamefont{and}
  \bibinfo{author}{\bibfnamefont{V.~P.} \bibnamefont{K}},
  \bibinfo{journal}{JHEP} \textbf{\bibinfo{volume}{10}}, \bibinfo{pages}{240}
  (\bibinfo{year}{2021}), \eprint{2104.03291}.

\bibitem[{\citenamefont{Brdar et~al.}(2021)\citenamefont{Brdar, Greljo, Kopp,
  and Opferkuch}}]{Brdar:2020quo}
\bibinfo{author}{\bibfnamefont{V.}~\bibnamefont{Brdar}},
  \bibinfo{author}{\bibfnamefont{A.}~\bibnamefont{Greljo}},
  \bibinfo{author}{\bibfnamefont{J.}~\bibnamefont{Kopp}}, \bibnamefont{and}
  \bibinfo{author}{\bibfnamefont{T.}~\bibnamefont{Opferkuch}},
  \bibinfo{journal}{JCAP} \textbf{\bibinfo{volume}{01}}, \bibinfo{pages}{039}
  (\bibinfo{year}{2021}), \eprint{2007.15563}.

\bibitem[{\citenamefont{Khan}(2023{\natexlab{a}})}]{Khan:2022bel}
\bibinfo{author}{\bibfnamefont{A.~N.} \bibnamefont{Khan}},
  \bibinfo{journal}{Phys. Lett. B} \textbf{\bibinfo{volume}{837}},
  \bibinfo{pages}{137650} (\bibinfo{year}{2023}{\natexlab{a}}),
  \eprint{2208.02144}.

\bibitem[{\citenamefont{Khan}(2023{\natexlab{b}})}]{Khan:2022akj}
\bibinfo{author}{\bibfnamefont{A.~N.} \bibnamefont{Khan}},
  \bibinfo{journal}{Nucl. Phys. B} \textbf{\bibinfo{volume}{986}},
  \bibinfo{pages}{116064} (\bibinfo{year}{2023}{\natexlab{b}}),
  \eprint{2201.10578}.

\bibitem[{\citenamefont{Jeong et~al.}(2021)\citenamefont{Jeong, Kim, and
  Youn}}]{Jeong:2021ivd}
\bibinfo{author}{\bibfnamefont{J.}~\bibnamefont{Jeong}},
  \bibinfo{author}{\bibfnamefont{J.~E.} \bibnamefont{Kim}}, \bibnamefont{and}
  \bibinfo{author}{\bibfnamefont{S.}~\bibnamefont{Youn}},
  \bibinfo{journal}{Int. J. Mod. Phys. A} \textbf{\bibinfo{volume}{36}},
  \bibinfo{pages}{2150182} (\bibinfo{year}{2021}), \eprint{2105.01842}.

\bibitem[{\citenamefont{Alok et~al.}(2023)\citenamefont{Alok, Singh~Chundawat,
  and Mandal}}]{Alok:2022pdn}
\bibinfo{author}{\bibfnamefont{A.~K.} \bibnamefont{Alok}},
  \bibinfo{author}{\bibfnamefont{N.~R.} \bibnamefont{Singh~Chundawat}},
  \bibnamefont{and} \bibinfo{author}{\bibfnamefont{A.}~\bibnamefont{Mandal}},
  \bibinfo{journal}{Phys. Lett. B} \textbf{\bibinfo{volume}{839}},
  \bibinfo{pages}{137791} (\bibinfo{year}{2023}), \eprint{2207.13034}.

\bibitem[{\citenamefont{Dror et~al.}(2021)\citenamefont{Dror, Elor, McGehee,
  and Yu}}]{Dror:2020czw}
\bibinfo{author}{\bibfnamefont{J.~A.} \bibnamefont{Dror}},
  \bibinfo{author}{\bibfnamefont{G.}~\bibnamefont{Elor}},
  \bibinfo{author}{\bibfnamefont{R.}~\bibnamefont{McGehee}}, \bibnamefont{and}
  \bibinfo{author}{\bibfnamefont{T.-T.} \bibnamefont{Yu}},
  \bibinfo{journal}{Phys. Rev. D} \textbf{\bibinfo{volume}{103}},
  \bibinfo{pages}{035001} (\bibinfo{year}{2021}), \bibinfo{note}{[Erratum:
  Phys.Rev.D 105, 119903 (2022)]}, \eprint{2011.01940}.

\bibitem[{\citenamefont{Agostini et~al.}(2017)}]{Borexino:2017fbd}
\bibinfo{author}{\bibfnamefont{M.}~\bibnamefont{Agostini}} \bibnamefont{et~al.}
  (\bibinfo{collaboration}{Borexino}), \bibinfo{journal}{Phys. Rev. D}
  \textbf{\bibinfo{volume}{96}}, \bibinfo{pages}{091103}
  (\bibinfo{year}{2017}), \eprint{1707.09355}.

\bibitem[{\citenamefont{Wong et~al.}(2007)}]{TEXONO:2006xds}
\bibinfo{author}{\bibfnamefont{H.~T.} \bibnamefont{Wong}} \bibnamefont{et~al.}
  (\bibinfo{collaboration}{TEXONO}), \bibinfo{journal}{Phys. Rev. D}
  \textbf{\bibinfo{volume}{75}}, \bibinfo{pages}{012001}
  (\bibinfo{year}{2007}), \eprint{hep-ex/0605006}.

\bibitem[{\citenamefont{Liu et~al.}(2004)}]{Super-Kamiokande:2004wqk}
\bibinfo{author}{\bibfnamefont{D.~W.} \bibnamefont{Liu}} \bibnamefont{et~al.}
  (\bibinfo{collaboration}{Super-Kamiokande}), \bibinfo{journal}{Phys. Rev.
  Lett.} \textbf{\bibinfo{volume}{93}}, \bibinfo{pages}{021802}
  (\bibinfo{year}{2004}), \eprint{hep-ex/0402015}.

\bibitem[{\citenamefont{Miller~Bertolami}(2014)}]{MillerBertolami:2014oki}
\bibinfo{author}{\bibfnamefont{M.~M.} \bibnamefont{Miller~Bertolami}},
  \bibinfo{journal}{Astron. Astrophys.} \textbf{\bibinfo{volume}{562}},
  \bibinfo{pages}{A123} (\bibinfo{year}{2014}), \eprint{1407.1404}.

\bibitem[{\citenamefont{Keus et~al.}(2014{\natexlab{a}})\citenamefont{Keus,
  King, Moretti, and Sokolowska}}]{Keus:2014jha}
\bibinfo{author}{\bibfnamefont{V.}~\bibnamefont{Keus}},
  \bibinfo{author}{\bibfnamefont{S.~F.} \bibnamefont{King}},
  \bibinfo{author}{\bibfnamefont{S.}~\bibnamefont{Moretti}}, \bibnamefont{and}
  \bibinfo{author}{\bibfnamefont{D.}~\bibnamefont{Sokolowska}},
  \bibinfo{journal}{JHEP} \textbf{\bibinfo{volume}{11}}, \bibinfo{pages}{016}
  (\bibinfo{year}{2014}{\natexlab{a}}), \eprint{1407.7859}.

\bibitem[{\citenamefont{Keus et~al.}(2014{\natexlab{b}})\citenamefont{Keus,
  King, and Moretti}}]{Keus:2014isa}
\bibinfo{author}{\bibfnamefont{V.}~\bibnamefont{Keus}},
  \bibinfo{author}{\bibfnamefont{S.~F.} \bibnamefont{King}}, \bibnamefont{and}
  \bibinfo{author}{\bibfnamefont{S.}~\bibnamefont{Moretti}},
  \bibinfo{journal}{Phys. Rev. D} \textbf{\bibinfo{volume}{90}},
  \bibinfo{pages}{075015} (\bibinfo{year}{2014}{\natexlab{b}}),
  \eprint{1408.0796}.

\bibitem[{\citenamefont{Cao et~al.}(2007)\citenamefont{Cao, Ma, and
  Rajasekaran}}]{Cao:2007rm}
\bibinfo{author}{\bibfnamefont{Q.-H.} \bibnamefont{Cao}},
  \bibinfo{author}{\bibfnamefont{E.}~\bibnamefont{Ma}}, \bibnamefont{and}
  \bibinfo{author}{\bibfnamefont{G.}~\bibnamefont{Rajasekaran}},
  \bibinfo{journal}{Phys. Rev. D} \textbf{\bibinfo{volume}{76}},
  \bibinfo{pages}{095011} (\bibinfo{year}{2007}), \eprint{0708.2939}.

\bibitem[{\citenamefont{Lundstrom et~al.}(2009)\citenamefont{Lundstrom,
  Gustafsson, and Edsjo}}]{Lundstrom:2008ai}
\bibinfo{author}{\bibfnamefont{E.}~\bibnamefont{Lundstrom}},
  \bibinfo{author}{\bibfnamefont{M.}~\bibnamefont{Gustafsson}},
  \bibnamefont{and} \bibinfo{author}{\bibfnamefont{J.}~\bibnamefont{Edsjo}},
  \bibinfo{journal}{Phys. Rev. D} \textbf{\bibinfo{volume}{79}},
  \bibinfo{pages}{035013} (\bibinfo{year}{2009}), \eprint{0810.3924}.

\bibitem[{\citenamefont{Cirelli et~al.}(2006)\citenamefont{Cirelli, Fornengo,
  and Strumia}}]{Cirelli:2005uq}
\bibinfo{author}{\bibfnamefont{M.}~\bibnamefont{Cirelli}},
  \bibinfo{author}{\bibfnamefont{N.}~\bibnamefont{Fornengo}}, \bibnamefont{and}
  \bibinfo{author}{\bibfnamefont{A.}~\bibnamefont{Strumia}},
  \bibinfo{journal}{Nucl. Phys. B} \textbf{\bibinfo{volume}{753}},
  \bibinfo{pages}{178} (\bibinfo{year}{2006}), \eprint{hep-ph/0512090}.

\bibitem[{\citenamefont{Xing and Zhou}(2011)}]{Xing:2011zza}
\bibinfo{author}{\bibfnamefont{Z.-z.} \bibnamefont{Xing}} \bibnamefont{and}
  \bibinfo{author}{\bibfnamefont{S.}~\bibnamefont{Zhou}},
  \emph{\bibinfo{title}{{Neutrinos in particle physics, astronomy and
  cosmology}}} (\bibinfo{year}{2011}), ISBN \bibinfo{isbn}{978-3-642-17559-6,
  978-7-308-08024-8}.

\bibitem[{\citenamefont{Babu et~al.}(2020)\citenamefont{Babu, Jana, and
  Lindner}}]{Babu:2020ivd}
\bibinfo{author}{\bibfnamefont{K.~S.} \bibnamefont{Babu}},
  \bibinfo{author}{\bibfnamefont{S.}~\bibnamefont{Jana}}, \bibnamefont{and}
  \bibinfo{author}{\bibfnamefont{M.}~\bibnamefont{Lindner}},
  \bibinfo{journal}{JHEP} \textbf{\bibinfo{volume}{10}}, \bibinfo{pages}{040}
  (\bibinfo{year}{2020}), \eprint{2007.04291}.

\bibitem[{\citenamefont{Chen et~al.}(2021)\citenamefont{Chen, Dutta~Banik, and
  Liu}}]{Chen:2020ark}
\bibinfo{author}{\bibfnamefont{S.-L.} \bibnamefont{Chen}},
  \bibinfo{author}{\bibfnamefont{A.}~\bibnamefont{Dutta~Banik}},
  \bibnamefont{and} \bibinfo{author}{\bibfnamefont{Z.-K.} \bibnamefont{Liu}},
  \bibinfo{journal}{Nucl. Phys. B} \textbf{\bibinfo{volume}{966}},
  \bibinfo{pages}{115394} (\bibinfo{year}{2021}), \eprint{2011.13551}.

\bibitem[{\citenamefont{Lineros and Pierre}(2020)}]{Lineros:2020eit}
\bibinfo{author}{\bibfnamefont{R.~A.} \bibnamefont{Lineros}} \bibnamefont{and}
  \bibinfo{author}{\bibfnamefont{M.}~\bibnamefont{Pierre}},
  \bibinfo{journal}{JHEP} \textbf{\bibinfo{volume}{21}}, \bibinfo{pages}{072}
  (\bibinfo{year}{2020}), \eprint{2011.08195}.

\bibitem[{\citenamefont{\'Avila et~al.}(2020)\citenamefont{\'Avila, De~Romeri,
  Duarte, and Valle}}]{Avila:2019hhv}
\bibinfo{author}{\bibfnamefont{I.~M.} \bibnamefont{\'Avila}},
  \bibinfo{author}{\bibfnamefont{V.}~\bibnamefont{De~Romeri}},
  \bibinfo{author}{\bibfnamefont{L.}~\bibnamefont{Duarte}}, \bibnamefont{and}
  \bibinfo{author}{\bibfnamefont{J.~W.~F.} \bibnamefont{Valle}},
  \bibinfo{journal}{Eur. Phys. J. C} \textbf{\bibinfo{volume}{80}},
  \bibinfo{pages}{908} (\bibinfo{year}{2020}), \eprint{1910.08422}.

\bibitem[{\citenamefont{Lopez~Honorez et~al.}(2007)\citenamefont{Lopez~Honorez,
  Nezri, Oliver, and Tytgat}}]{LopezHonorez:2006gr}
\bibinfo{author}{\bibfnamefont{L.}~\bibnamefont{Lopez~Honorez}},
  \bibinfo{author}{\bibfnamefont{E.}~\bibnamefont{Nezri}},
  \bibinfo{author}{\bibfnamefont{J.~F.} \bibnamefont{Oliver}},
  \bibnamefont{and} \bibinfo{author}{\bibfnamefont{M.~H.~G.}
  \bibnamefont{Tytgat}}, \bibinfo{journal}{JCAP} \textbf{\bibinfo{volume}{02}},
  \bibinfo{pages}{028} (\bibinfo{year}{2007}), \eprint{hep-ph/0612275}.

\bibitem[{\citenamefont{Barbieri et~al.}(2006)\citenamefont{Barbieri, Hall, and
  Rychkov}}]{Barbieri:2006dq}
\bibinfo{author}{\bibfnamefont{R.}~\bibnamefont{Barbieri}},
  \bibinfo{author}{\bibfnamefont{L.~J.} \bibnamefont{Hall}}, \bibnamefont{and}
  \bibinfo{author}{\bibfnamefont{V.~S.} \bibnamefont{Rychkov}},
  \bibinfo{journal}{Phys. Rev. D} \textbf{\bibinfo{volume}{74}},
  \bibinfo{pages}{015007} (\bibinfo{year}{2006}), \eprint{hep-ph/0603188}.

\bibitem[{\citenamefont{Dolle and Su}(2009)}]{Dolle:2009fn}
\bibinfo{author}{\bibfnamefont{E.~M.} \bibnamefont{Dolle}} \bibnamefont{and}
  \bibinfo{author}{\bibfnamefont{S.}~\bibnamefont{Su}}, \bibinfo{journal}{Phys.
  Rev. D} \textbf{\bibinfo{volume}{80}}, \bibinfo{pages}{055012}
  (\bibinfo{year}{2009}), \eprint{0906.1609}.

\bibitem[{\citenamefont{Ellis et~al.}(2000)\citenamefont{Ellis, Ferstl, and
  Olive}}]{Ellis:2000ds}
\bibinfo{author}{\bibfnamefont{J.~R.} \bibnamefont{Ellis}},
  \bibinfo{author}{\bibfnamefont{A.}~\bibnamefont{Ferstl}}, \bibnamefont{and}
  \bibinfo{author}{\bibfnamefont{K.~A.} \bibnamefont{Olive}},
  \bibinfo{journal}{Phys. Lett. B} \textbf{\bibinfo{volume}{481}},
  \bibinfo{pages}{304} (\bibinfo{year}{2000}), \eprint{hep-ph/0001005}.

\bibitem[{\citenamefont{Semenov}(1996)}]{Semenov:1996es}
\bibinfo{author}{\bibfnamefont{A.~V.} \bibnamefont{Semenov}}
  (\bibinfo{year}{1996}), \eprint{hep-ph/9608488}.

\bibitem[{\citenamefont{Pukhov et~al.}(1999)\citenamefont{Pukhov, Boos,
  Dubinin, Edneral, Ilyin, Kovalenko, Kryukov, Savrin, Shichanin, and
  Semenov}}]{Pukhov:1999gg}
\bibinfo{author}{\bibfnamefont{A.}~\bibnamefont{Pukhov}},
  \bibinfo{author}{\bibfnamefont{E.}~\bibnamefont{Boos}},
  \bibinfo{author}{\bibfnamefont{M.}~\bibnamefont{Dubinin}},
  \bibinfo{author}{\bibfnamefont{V.}~\bibnamefont{Edneral}},
  \bibinfo{author}{\bibfnamefont{V.}~\bibnamefont{Ilyin}},
  \bibinfo{author}{\bibfnamefont{D.}~\bibnamefont{Kovalenko}},
  \bibinfo{author}{\bibfnamefont{A.}~\bibnamefont{Kryukov}},
  \bibinfo{author}{\bibfnamefont{V.}~\bibnamefont{Savrin}},
  \bibinfo{author}{\bibfnamefont{S.}~\bibnamefont{Shichanin}},
  \bibnamefont{and} \bibinfo{author}{\bibfnamefont{A.}~\bibnamefont{Semenov}}
  (\bibinfo{year}{1999}), \eprint{hep-ph/9908288}.

\bibitem[{\citenamefont{Belanger et~al.}(2007)\citenamefont{Belanger, Boudjema,
  Pukhov, and Semenov}}]{Belanger:2006is}
\bibinfo{author}{\bibfnamefont{G.}~\bibnamefont{Belanger}},
  \bibinfo{author}{\bibfnamefont{F.}~\bibnamefont{Boudjema}},
  \bibinfo{author}{\bibfnamefont{A.}~\bibnamefont{Pukhov}}, \bibnamefont{and}
  \bibinfo{author}{\bibfnamefont{A.}~\bibnamefont{Semenov}},
  \bibinfo{journal}{Comput. Phys. Commun.} \textbf{\bibinfo{volume}{176}},
  \bibinfo{pages}{367} (\bibinfo{year}{2007}), \eprint{hep-ph/0607059}.

\bibitem[{\citenamefont{Belanger et~al.}(2009)\citenamefont{Belanger, Boudjema,
  Pukhov, and Semenov}}]{Belanger:2008sj}
\bibinfo{author}{\bibfnamefont{G.}~\bibnamefont{Belanger}},
  \bibinfo{author}{\bibfnamefont{F.}~\bibnamefont{Boudjema}},
  \bibinfo{author}{\bibfnamefont{A.}~\bibnamefont{Pukhov}}, \bibnamefont{and}
  \bibinfo{author}{\bibfnamefont{A.}~\bibnamefont{Semenov}},
  \bibinfo{journal}{Comput. Phys. Commun.} \textbf{\bibinfo{volume}{180}},
  \bibinfo{pages}{747} (\bibinfo{year}{2009}), \eprint{0803.2360}.

\bibitem[{\citenamefont{Aghanim et~al.}(2018)}]{Aghanim:2018eyx}
\bibinfo{author}{\bibfnamefont{N.}~\bibnamefont{Aghanim}} \bibnamefont{et~al.}
  (\bibinfo{collaboration}{Planck}) (\bibinfo{year}{2018}),
  \eprint{1807.06209}.

\bibitem[{\citenamefont{Aalbers et~al.}(2022)}]{LZ:2022ufs}
\bibinfo{author}{\bibfnamefont{J.}~\bibnamefont{Aalbers}} \bibnamefont{et~al.}
  (\bibinfo{collaboration}{LZ}) (\bibinfo{year}{2022}), \eprint{2207.03764}.

\bibitem[{\citenamefont{Kashiwase and Suematsu}(2013)}]{Kashiwase:2013uy}
\bibinfo{author}{\bibfnamefont{S.}~\bibnamefont{Kashiwase}} \bibnamefont{and}
  \bibinfo{author}{\bibfnamefont{D.}~\bibnamefont{Suematsu}},
  \bibinfo{journal}{Eur. Phys. J. C} \textbf{\bibinfo{volume}{73}},
  \bibinfo{pages}{2484} (\bibinfo{year}{2013}), \eprint{1301.2087}.

\bibitem[{\citenamefont{Ho and Tandean}(2013)}]{Ho:2013hia}
\bibinfo{author}{\bibfnamefont{S.-Y.} \bibnamefont{Ho}} \bibnamefont{and}
  \bibinfo{author}{\bibfnamefont{J.}~\bibnamefont{Tandean}},
  \bibinfo{journal}{Phys. Rev. D} \textbf{\bibinfo{volume}{87}},
  \bibinfo{pages}{095015} (\bibinfo{year}{2013}), \eprint{1303.5700}.

\bibitem[{\citenamefont{Singirala}(2017)}]{Singirala:2016kam}
\bibinfo{author}{\bibfnamefont{S.}~\bibnamefont{Singirala}},
  \bibinfo{journal}{Chin. Phys. C} \textbf{\bibinfo{volume}{41}},
  \bibinfo{pages}{043102} (\bibinfo{year}{2017}), \eprint{1607.03309}.

\bibitem[{\citenamefont{Esteban et~al.}(2020)\citenamefont{Esteban,
  Gonzalez-Garcia, Maltoni, Schwetz, and Zhou}}]{Esteban:2020cvm}
\bibinfo{author}{\bibfnamefont{I.}~\bibnamefont{Esteban}},
  \bibinfo{author}{\bibfnamefont{M.~C.} \bibnamefont{Gonzalez-Garcia}},
  \bibinfo{author}{\bibfnamefont{M.}~\bibnamefont{Maltoni}},
  \bibinfo{author}{\bibfnamefont{T.}~\bibnamefont{Schwetz}}, \bibnamefont{and}
  \bibinfo{author}{\bibfnamefont{A.}~\bibnamefont{Zhou}},
  \bibinfo{journal}{JHEP} \textbf{\bibinfo{volume}{09}}, \bibinfo{pages}{178}
  (\bibinfo{year}{2020}), \eprint{2007.14792}.

\bibitem[{\citenamefont{Roy~Choudhury and
  Choubey}(2018)}]{RoyChoudhury:2018gay}
\bibinfo{author}{\bibfnamefont{S.}~\bibnamefont{Roy~Choudhury}}
  \bibnamefont{and} \bibinfo{author}{\bibfnamefont{S.}~\bibnamefont{Choubey}},
  \bibinfo{journal}{JCAP} \textbf{\bibinfo{volume}{09}}, \bibinfo{pages}{017}
  (\bibinfo{year}{2018}), \eprint{1806.10832}.

\bibitem[{\citenamefont{Giunti et~al.}(2016)\citenamefont{Giunti, Kouzakov, Li,
  Lokhov, Studenikin, and Zhou}}]{Giunti:2015gga}
\bibinfo{author}{\bibfnamefont{C.}~\bibnamefont{Giunti}},
  \bibinfo{author}{\bibfnamefont{K.~A.} \bibnamefont{Kouzakov}},
  \bibinfo{author}{\bibfnamefont{Y.-F.} \bibnamefont{Li}},
  \bibinfo{author}{\bibfnamefont{A.~V.} \bibnamefont{Lokhov}},
  \bibinfo{author}{\bibfnamefont{A.~I.} \bibnamefont{Studenikin}},
  \bibnamefont{and} \bibinfo{author}{\bibfnamefont{S.}~\bibnamefont{Zhou}},
  \bibinfo{journal}{Annalen Phys.} \textbf{\bibinfo{volume}{528}},
  \bibinfo{pages}{198} (\bibinfo{year}{2016}), \eprint{1506.05387}.

\bibitem[{\citenamefont{Aprile et~al.}(2020{\natexlab{b}})}]{XENON:2020iwh}
\bibinfo{author}{\bibfnamefont{E.}~\bibnamefont{Aprile}} \bibnamefont{et~al.}
  (\bibinfo{collaboration}{XENON}), \bibinfo{journal}{Eur. Phys. J. C}
  \textbf{\bibinfo{volume}{80}}, \bibinfo{pages}{785}
  (\bibinfo{year}{2020}{\natexlab{b}}), \eprint{2003.03825}.

\bibitem[{\citenamefont{Bahcall and Pena-Garay}(2004)}]{Bahcall:2004mz}
\bibinfo{author}{\bibfnamefont{J.~N.} \bibnamefont{Bahcall}} \bibnamefont{and}
  \bibinfo{author}{\bibfnamefont{C.}~\bibnamefont{Pena-Garay}},
  \bibinfo{journal}{New J. Phys.} \textbf{\bibinfo{volume}{6}},
  \bibinfo{pages}{63} (\bibinfo{year}{2004}), \eprint{hep-ph/0404061}.

\bibitem[{\citenamefont{Lopes and Turck-Chi\`eze}(2013)}]{Lopes:2013nfa}
\bibinfo{author}{\bibfnamefont{I.}~\bibnamefont{Lopes}} \bibnamefont{and}
  \bibinfo{author}{\bibfnamefont{S.}~\bibnamefont{Turck-Chi\`eze}},
  \bibinfo{journal}{Astrophys. J.} \textbf{\bibinfo{volume}{765}},
  \bibinfo{pages}{14} (\bibinfo{year}{2013}), \eprint{1302.2791}.

\bibitem[{\citenamefont{Zyla et~al.}(2020)}]{ParticleDataGroup:2020ssz}
\bibinfo{author}{\bibfnamefont{P.~A.} \bibnamefont{Zyla}} \bibnamefont{et~al.}
  (\bibinfo{collaboration}{Particle Data Group}), \bibinfo{journal}{PTEP}
  \textbf{\bibinfo{volume}{2020}}, \bibinfo{pages}{083C01}
  (\bibinfo{year}{2020}).

\end{thebibliography}
\end{document}